\documentclass[a4paper,fleqn,usenatbib,useAMS]{mnras}

\usepackage{graphicx}	
\usepackage{amsmath}	
\usepackage{amssymb}	
\usepackage{multicol}        
\usepackage{bm}		
\usepackage{pdflscape}	
\usepackage{natbib}	

\newcommand{\paperI}[1]{\textcolor{black}{Paper~I}}
\newcommand{\ppxf}[1]{\textsc{ppxf}}

\defcitealias{Thomas_2011}{TMJ}

\usepackage[T1]{fontenc}
\usepackage{ae,aecompl}

\usepackage{newtxtext,newtxmath}

\title[BUDDI-MaNGA II: SFHs of S0s]{ BUDDI-MaNGA II: The Star-Formation Histories of Bulges and Discs of S0s}

\author[E. J. Johnston \textit{et al}]{Evelyn J. Johnston$^{1}$\thanks{Contact e-mail: \href{mailto:evelyn.johnston@mail.udp.cl}{evelyn.johnston@mail.udp.cl}}, Boris H\"au\ss ler$^{2}$,  Keerthana Jegatheesan$^{1}$,  
\newauthor  Amelia Fraser-McKelvie$^{3,4}$, Lodovico Coccato$^{5}$, Ariana Cortesi$^{6,7}$, Yara Jaff\'e$^{8}$,
\newauthor  Gaspar Galaz$^{9}$, Marcelo Mora$^{9}$ \& Yasna Ordenes-Brice\~no$^{9}$.
\\
$^{1}$ N\'ucleo de Astronom\'ia, Facultad de Ingenier\'ia y Ciencias, Universidad Diego Portales, Av. Ej\'ercito Libertador 441, Santiago, Chile\\
$^{2}$ European Southern Observatory, Alonso de Cordova 3107, Vitacura, Santiago, Chile\\
 $^{3}$ International Centre for Radio Astronomy Research, The University of Western Australia, 35 Stirling Hwy, 6009 Crawley, WA, Australia \\
 $^{4}$ ARC Centre of Excellence for All Sky Astrophysics in 3 Dimensions (ASTRO 3D) \\
 $^{5}$ European Southern Observatory, Karl-Schwarzchild-str., 2, D-85748 Garching b. Munchen, Germany\\
$^{6}$ Centro Brasileiro de Pesquisas F\'isicas, Rua Dr. Xavier Sigaud 150, CEP 22290-180, Rio de Janeiro, RJ, Brazil\\ 
$^{7}$ Observatorio de Valongo, Universidade Federal do Rio de Janeiro, Ladeira do Pedro Antonio, 43, Centro, Rio de Janeiro - RJ 20080-090, Brazil \\
$^{8}$ Instituto de Física y Astronom\'ia, Universidad de Valpara\'iso, Gran Bretana 1111, Valpara\'iso, Chile\\
$^{9}$ Instituto de Astrofisica, Pontificia Universidad Cat\'olica de Chile, Av.~Vicu\~na Mackenna 4860, 7820436 Macul, Santiago, Chile.}

\pubyear{2022}

\begin{document}
\label{firstpage}
\pagerange{\pageref{firstpage}--\pageref{lastpage}}
\maketitle

\begin{abstract}
Many  processes have been proposed to explain the quenching of star formation in spiral galaxies and their transformation into S0s. These processes  affect the bulge and disc in different ways, and so by isolating the bulge and disc spectra, we can look for these characteristic signatures. In this work, we used  \textsc{buddi}  to cleanly extract the spectra of the bulges and discs of 78 S0 galaxies in the MaNGA Survey. We compared the luminosity and mass weighted stellar populations of the bulges and discs, finding that bulges are generally older and more metal rich than their discs. When considering the mass and environment of each galaxy, we found that the galaxy stellar mass  plays a more significant role on the formation of the bulges. Bulges in galaxies with masses $\geq10^{10}M_\odot$ built up the majority of their mass rapidly early in their lifetimes, while those in lower mass galaxies formed over more extended timescales and more recently. No clear difference was found in the formation or quenching processes of the discs as a function of galaxy environment. We conclude that more massive S0 galaxies formed through an inside-out scenario, where the bulge formed first and evolved passively while the disc underwent a more extended period of star formation. In lower mass S0s,  the bulges and discs either formed together from the same material, or through an outside-in scenario. Our results therefore imply multiple formation mechanisms for S0 galaxies, the pathway of which is chiefly determined by a galaxy's current stellar mass.

\end{abstract}




\section{Introduction}
Lenticular, or S0, galaxies are generally considered to be one possible endpoint in the evolution of spirals, in which the bulge+disc morphology has remained while the star formation was truncated and the spiral arms faded \citep[e.g. ][]{Bedregal_2006, Moran_2007, Laurikainen_2010, Cappellari_2011, Prochaska_2011, Kormendy_2012}. This evolution has been found to have a connection to both the environment and redshift. For example, S0s have been found in higher frequencies in regions with higher densities, such as the inner regions of clusters, and as you move to lower redshift, while the fraction of spirals decreases proportionally \citep{Dressler_1980, Dressler_1997, Fasano_2000}. However, the transformation from a spiral to an S0 is not simple, and is further hindered by the potential for misclassification, where a face-on S0 can easily be mistaken for an elliptical, and vice versa \citep[e.g.][]{vandenBergh_2009}. For this reason, many studies combine these two galaxy morphologies together under the term Early-Type Galaxy (ETG), and study their evolution together despite the clear differences in their morphologies and kinematics.

Many theories have been proposed in the literature to explain the formation of S0s, many of which focus on the processes that lead to the truncation of the star formation in the disc. For example, many theories focus on the gas being stripped out via ram-pressure stripping \citep{Gunn_1972}, harrassment \citep{Moore_1998} or starvation/strangulation \citep{Larson_1980}, or used up in a rapid star formation event following a minor merger \citep{Ponman_1994, Arnold_2011} or a series of minor mergers \citep{Bekki_2011}. Major mergers have also been shown to create S0s in simulations by first creating a bulge around which a disc forms later on \citep{Spitzer_1951, Tapia_2017, Diaz_2018, Eliche_2018, Mendez_2018}, while \citet{Deeley_2021} used simulations from IllustrisTNG to show evidence of two main formation pathways for S0s- gas stripping during group or cluster infall, and major mergers.  

While all of these processes are triggered by an external influence, many S0s are now known to reside in isolated environments, raising the question of whether they formed during an isolated interaction or through internal processes. For example, could S0s simply be remnants of spirals after the gas reservoirs become exhausted and the spirals arms have faded, as proposed by \citet{Eliche_2013}? Or perhaps they are the result of disc instabilities and fragmentation, as suggested by \citet{Saha_2018}.

Each of these processes would affect the bulge and disc of the galaxy in different ways, leaving characteristic signatures in their spectra that can be used to better understand how the star formation was truncated and how the galaxy transformed. For example, gas stripping processes may produce stellar population gradients across the galaxy where the star formation becomes truncated first in the outskirts while continuing for some time in the inner regions \citep[e.g.][]{Pfeffer_2022}. On the other hand, minor mergers might trigger a short episode of enhanced star formation across the entire galaxy, using up all the gas at around the same time and producing flatter age gradients \citep{Mehlert_2003, Ogando_2005,Kuntschner_2010, Eigenthaler_2013}. Additionally, galaxies that experienced recent minor mergers (i.e. within 2-3~Gyrs) may still show evidence of these interactions in the form of shells and tidal tails \citep{Malin_1983,Wilkinson_2000, Duc_2015, Alabi_2020}, some of which may only be possible to detect after subtracting a model of the smoothed light profile of the galaxy \citep[e.g.][]{Rampazzo_2020, Johnston_2021}. With spatially-resolved spectroscopy, we can detect variations in the stellar populations and star formation histories across galaxies, and attempts have been made to extract the spectra from the bulges and discs. 

One common approach is spectroscopic bulge--disc decomposition, where the light profile of the galaxy is used to separate the light of clear visual components. For example, \citet{Johnston_2012,Johnston_2014} modelled the 1-dimensional light profile of S0 galaxies in the Fornax and Virgo Clusters at each wavelength using long-slit spectra aligned with the major axes of each galaxy. They used a S\'ersic and exponential profiles to model the bulge and disc components respectively. From the luminosity-weighted stellar populations analysis of these spectra, they found that in general, the bulges of these galaxies were younger and more metal rich than the discs, and concluded that during the truncation of the star formation within the disc, some of the gas was funnelled into the inner region of the galaxy where, later, it triggered a final episode of star formation within the bulge region.  A similar approach to extract the bulge and disc spectra in long-slit spectra was proposed by \citet{Silchenko_2012}, who also found evidence of systematically younger and more metal-rich stellar populations within the bulges.

However, with long-slit spectra alone, it was impossible to determine the spatial distribution of these young stellar populations, i.e. whether they were distributed throughout the bulge or were instead concentrated in the centre of the disc, and were simply attributed to the bulge due to the constraint of modelling the disc as an exponential profile. To resolve this issue reliably, both the spatial and spectroscopic information must be used. The introduction of Integral Field Unit (IFU) spectrographs with wide fields of view and high spatial resolution, such as the Multi-Unit Spectroscopic Explorer \citep[MUSE,][]{Bacon_2010}, and surveys with these instruments, like the Mapping of Nearby Galaxies at APO \citep[MaNGA;][]{Bundy_2015} Survey, the Sydney-AAO Multi-object Integral field Galaxy Survey \citep[SAMI,][]{Bryant_2015} and the Calar Alto Legacy Integral Field Area  \citep[CALIFA,][]{Sanchez_2012} survey, this information is now available. 

One way that studies have attempted to disentangle the spectra of the bulges and discs is to use the decomposition of imaging data to determine regions within the galaxy that are dominated by light from the bulge and disc. \citet{Fraser_2018b} used this idea on a sample of 279 S0 galaxies in the MaNGA survey, and found that the bulges and discs are co-evolving. When comparing the luminosity-weighted stellar populations of the bulges and discs, they found that in higher mass galaxies, with masses $>10^{10}M_\odot$, the bulges are generally older than their discs, while in lower mass galaxies, the fraction of bulges with younger stellar populations increases. They concluded that the higher mass galaxies have undergone morphological or inside-out quenching of their star formation, while the lower mass galaxies may be undergoing a process of bulge rejuvenation or disc fading. \citet{Barsanti_2021} used a similar technique on a sample of S0s in the SAMI galaxy survey to measure the mass-weighted stellar populations of the bulges and discs. They again found a mix of bulge and disc ages, with 23\%, 34\% and 43\% of their sample containing bulges with older, younger and similar ages, respectively, relative to their discs. They also noticed that redder bulges tended to be more metal-rich than their discs, and concluded that the redder colour in bulges is due to their enhanced metallicity relative to their disks as opposed to being dependent on the ages of the stellar populations present. 

While these studies use the fits to imaging data to determine the regions within the galaxy where the bulge or disc dominates the light, a low level of contamination from other components is often present. This contamination can significantly affect the analysis. For example, in the case of an old quiescent bulge surrounded by a younger star-forming disc, the bulge will dominate the light at the centre of the galaxy, but the contamination of the young stars within the disc may artificially bias the age of the bulge to younger ages, particularly when considering the luminosity-weighted ages. In order to reduce this contamination, there have been several studies applying the light profile fits to the IFU data in order to cleanly separate the spectra of each component.

One of the first works to present this idea for IFU data was \citet{Johnston_2017}, who presented Bulge--Disc Decomposition of IFU data (\textsc{buddi}). \textsc{buddi} is a wrapper for \textsc{GalfitM} \citep{Haeussler_2013}, a modified form of \textsc{Galfit} \citep{Peng_2002,Peng_2010} that can model multi-waveband images simultaneously, thus using information from the entire dataset to fit each image. 
\textsc{buddi} extends this idea to modelling the 2-dimensional light profile of the galaxy in every image slice of an IFU datacube to create wavelength-dependent models of each component, from which their spectra could be cleanly extracted. \citet{Johnston_2021} used this code to decompose a sample of 8 S0 galaxies observed with MUSE, separating the light into bulge, disc and lens components in order to study the effect of environment on the formation of these galaxies. They found no clear trend in the luminosity-weighted ages of the bulges and discs, but the mass-weighted ages showed the bulges to be generally older than their discs. They also found that the bulges tend to have similar or higher metallicities than the discs, and that the $\alpha$-enhancement of the bulges and discs are connected. They concluded that the majority of the mass in those galaxies was built up early in the lifetime of the galaxy, where the bulges and discs formed from the same material and the star formation continued in the disc long after it was truncated in the bulge. When considering the environment, they also found that the field S0s showed generally younger stellar populations and more asymmetric features than the cluster galaxies, indicating that the star formation in these galaxies was more likely to have been truncated through minor mergers, while in cluster galaxies gas stripping processes were more dominant. However, the low number of galaxies in this sample prevented them from making any firm conclusions.

A similar technique to \textsc{buddi}, called C2D, was introduced by \citet{Mendez_2019a}, and was applied to a sample of 49 lenticular galaxies from the CALIFA in \citet{Mendez_2019b}. They found evidence of star formation only in the disc, not the bulge, and that star formation was distributed throughout the discs of these galaxies, as opposed to being concentrated in the outskirts. They concluded that the star formation in these galaxies was most likely quenched by morphological quenching. A follow-up study was carried out by \citet{Mendez_2021}, using a sample of 129 non-barred galaxies from the CALIFA survey with a range of morphologies, including pure ellipticals and those with clear bulge+disc structures. They found that while bulges formed early on in the lifetime of the galaxy and have experienced little evolution since, the early properties of the bulges are responsible for driving the evolution of the galaxy as a whole and of the discs that formed around them.

As these techniques develop and become more refined, the next natural step is to move on from small samples of galaxies, where individual anomalies can affect the results, to more statistical samples. Such samples can provide a more general overview of the evolution of galaxy components, and thus help us understand how common or rare different phenomena are. This work is the second in a series of papers as part of the BUDDI-MaNGA project, in which all suitable galaxies in the SDSS data release 15 \citep[DR15,][]{Aguado_2019} are modelled with \textsc{buddi} in order to extract the bulge and disc spectra, thus building up a more statistical sample in which the evolution of these components can be studied as a function of morphology and galaxy mass. \textcolor{black}{Johnston et al (submitted, hereafter \paperI\ )} gives an outline of the project, the data and methodology, and some tests on the robustness of the results derived with \textsc{buddi}. This work focusses on the first science case, comparing the properties of bulges and discs of S0s in order to better understand the processes that quenched the star formation and how they affect the stellar populations within the bulge and discs. 

The paper is laid out as follows: Section~\ref{sec:data} gives an overview of the data, the sample selection and the decomposition process, Section~\ref{sec:stellar_pops} presents the stellar populations analysis, Section~\ref{sec:discussion} discusses the results, and the conclusions are presented in Section~\ref{sec:conclusions}.

\section{Data, sample selection and decomposition}\label{sec:data}
\subsection{Data}\label{sec:sub_data}

This work uses data from SDSS Data Release 15 (hereafter DR15) of the  MaNGA Survey \citep{Bundy_2015}.  MaNGA is part of the Sloan Digital Sky Survey-IV \citep[SDSS-IV,][]{Blanton_2017}, and is an integral field spectroscopic survey using the BOSS spectrograph \citep{Smee_2013} on the 2.5 m SDSS telescope \citep{Gunn_2006} at the Apache Point Observatory. A total of 4824  galaxies are publicly available as part of DR15, covering redshifts in the range $0.01 < z < 0.15$ \citep{Yan_2016}. Since it was shown in \citet{Johnston_2017} that the 2-component fits to MaNGA data become unreliable for the 19, 37 and 61-fibre IFUs due to the smaller fields of view and fewer spaxels available, a sub-sample of $\sim1,900$ galaxies was selected for this study that were observed with the 91 and 127-fibre IFUs. Each galaxy is observed with a hexagonal IFU of different sizes, from the 19-fibre IFUs which have a 12\arcsec\ diameter up to the 127-fibre IFUs with a diameter of 32\arcsec\ \citep{Drory_2015}, and the IFU size is selected to provide coverage out to $\sim1.5R_e$ for $\sim66$~per cent of the total sample, and out to $\sim2.5R_e$ for the remaining galaxies. The final data has a continuous wavelength coverage between 3600 to 10300~\AA, with a spectral resolution of $R\sim1400$ at 4000~\AA\ to $R\sim2600$ at 9000~\AA\ \citep{Drory_2015}. Further details of the sample used in the BUDDI-MaNGA project can be found in \paperI\..

\subsection{Galaxy Decomposition with \textsc{buddi}}\label{sec:sub_decomp}
\paperI\. describes the decomposition of the datacubes for each galaxy with \textsc{buddi}, and a full description of how \textsc{buddi} works is described in \citet{Johnston_2017}. However, a short summary is included below.

The first step in the process was to measure and normalise the kinematics in the galaxy datacubes. This step is necessary since \textsc{Galfit} can only model symmetric structures, and so variations in the flux across a galaxy in individual image slices due to the position or shape of spectral features must be eliminated. The datacubes were binned using the Voronoi tessellation technique of \citet{Cappellari_2003} to reach a S/N of 50\AA$^{-1}$, and  the penalised Pixel Fitting software (\textsc{ppxf}) of \citet{Cappellari_2004} was used to measure the line-of-sight velocities and velocity dispersions in the binned spectra. This step used the MILES stellar library of \citet{Sanchez_2006}, which consisted of 156 template spectra in the age range of 1 to 17.78~Gyrs and with metallicities [M/H] between $-1.72$ to $+0.22$. Once the kinematics information was known, the spectra in each spaxel were shifted to match the line-of-sight velocity of the centre of the galaxy, defined as the brightest spaxel, and broadened to match the maximum velocity dispersion measured within the galaxy.

The datacube was then stacked to create a single broad-band image of the galaxy, which was modelled with \textsc{GalfitM}. For each galaxy, 2 fits were applied- a Single S\'ersic (SS) model and a S\'ersic+exponential model (SE). For both sets of fits, the initial parameters came from the fits to the $r$-band images in the MaNGA PyMorph DR15 Photometric Value Added Catalog \citep[MPP-VAC;][]{Fischer_2019}. Since this catalog includes a flag parameter to mark failed fits, usually due to irregular morphologies or contamination from foreground objects, only those galaxies without that failed flag were modelled with \textsc{buddi}. This constraint resulted in a total of $\sim1,700$ candidate galaxies for the SS and SE fits.

The next step involved rebinning the datacube along the wavelength direction into a series of narrow-band images, which were again modelled with \textsc{GalfitM} using the fit parameters from the previous step as the initial estimates. This step thus allows for further refinement of the fit parameters as a function of wavelength, which were then used in the next step for the fits to the individual images. In this case, the structural parameters were held fixed to the values derived from the fits to the narrow-band images, with only the magnitudes allowed to vary.

Finally, after modelling each image slice, the 1-dimensional spectra representing purely the light from each component are  created by plotting the flux as a function of wavelength. The fits to around 400 galaxies for both the SS and SE models failed, i.e. \textsc{GalfitM} failed to converge on a fit as the fit parameters exceeded the built in limits (see \paperI\. for more details). Of the remaining galaxies, the final sample of `good' fits were selected based on the structural parameters from the fits:
\begin{enumerate}
 \item $\Delta mag_r \leq 2.5$ (SE models only) 
 \item $0.25 \leq R_e \leq 50\arcsec$,
 \item $0.205 \leq n \leq 0.795$,
\end{enumerate}
where  $\Delta mag_r$ is the difference in the magnitudes over the wavelength range of the $r$-band between the S\'ersic and exponential components. Using these parameters, a further 100 and 400 fits were considered bad for the SS and SE models respectively. Consequently, of the $\sim1,700$ candidate galaxies, good spectra were derived for the SS model in 1,038 galaxies, and for 691 galaxies using the SE model. An example of the decomposition and extracted spectra can be found in Figure~1 of \paperI\..

It should be noted that in the current version of the fits, no further steps were taken to mask out other objects within the FOV, such as foreground stars or background/neighbouring galaxies, which may ultimately affect the final fits. However, this step is to be included in the next round of fits using the data released as part of SDSS Data Release 17  \citep[DR17,][]{Abdurrouf_2021}. 
Additionally, we have performed tests on the robustness of the application of BUDDI onto the MaNGA data, and we have investigated the dependency of the results from the adopted photometric model (e.g. using different parameters, radial photometric profiles, orders of the Chebychev polynomials). All these investigations have been presented and discussed in \paperI\.. Furthermore, interested readers can also find information for prior tests of the \textsc{buddi} fits to simulated datacubes in \citet{Johnston_2017,Johnston_2018} and \citet{Johnston_2021}, and a comprehensive analysis of the separation of the properties of bulge and disc components in large surveys can be found in \citet{Haeussler_2022}.

\subsection{S0 Sample Selection}\label{sec:sub_sample}

The focus of this work is to study the properties of S0s. The visual morphologies of the galaxies were taken from the MaNGA Deep Learning Morphology Value Added Catalogue \citep[MDLM-VAC,][]{Dominguez_2018}, which classified morphologies by using Deep learning algorithms to create models that were trained and tested on the SDSS-DR7 images. In addition to the Hubble T-Type, the MDLM-VAC contains information on Galaxy Zoo like attributes (edge-on, barred, projected pairs, bulge prominence, roundness and mergers), and an additional probability to distinguish between pure ellipticals from S0s (P$_{\text{S0}}$). Additionally, all images of the galaxies were checked by eye to confirm or refine the T-Type for greater reliability. 

\begin{figure}
 \includegraphics[width=\linewidth]{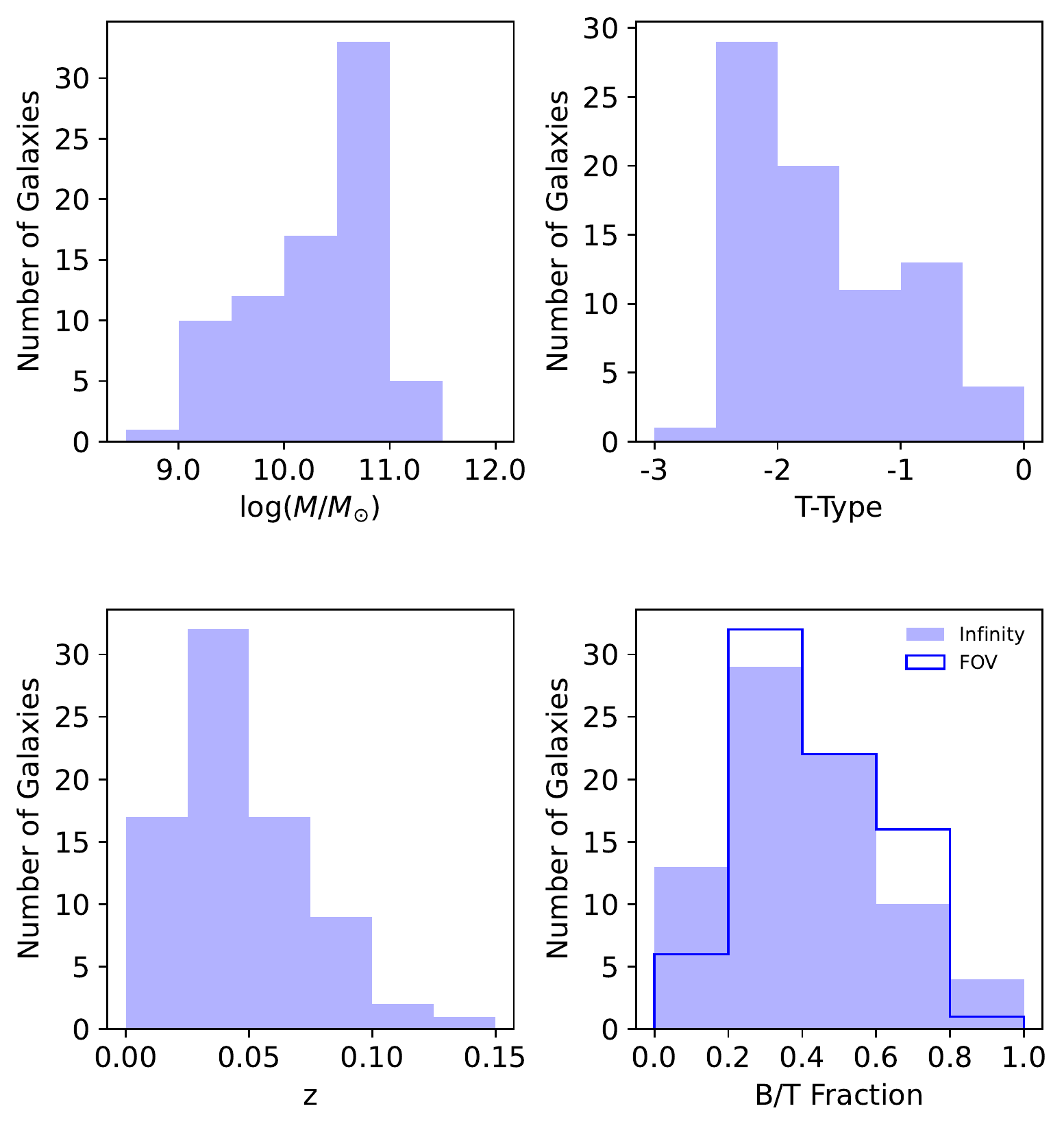}
 \caption{Overview of the mass, morphology, redshift and bulge-to-total light fraction distributions of the sample of S0 galaxies. In the bottom-right plot, the filled histogram represents the B/T ratios derived from the SE fits when the light profiles are integrated to infinity, while the histogram with solid lines and no filling shows the ratios within only the MaNGA FOV.}
 \label{fig:sample}
\end{figure}

The galaxies selected for this study were identified as those with early type visual morphologies (T-Type~$<0$) and with probabilities of being S0 P$_{\text{S0}}\ge0.5$ in the MDLM-VAC. This selection resulted in a final sample of 78 S0 galaxies. Figure~\ref{fig:sample} displays the distribution of these galaxies in terms of T-Type, stellar mass and redshift, showing that the majority of the sample have masses $\sim10^{10.5}M/M_\odot$ and lie at redshifts $z<0.10$, where these properties were taken from the NASA-Sloan Atlas catalog \citep[hereafter NSA\footnote{http://www.nsatlas.org/},][]{Blanton_2011}.  The B/T flux ratios are also given in this figure, as calculated from the models created by \textsc{buddi} both integrated to infinity and within the FOV of the datacube. The peak at low values for the B/T fraction thus reflects that the galaxies used in this sample were selected mainly based on visual morphology (i.e. lack of spiral arms) rather than the prominence of the bulge.

\begin{figure}
 \includegraphics[width=\linewidth]{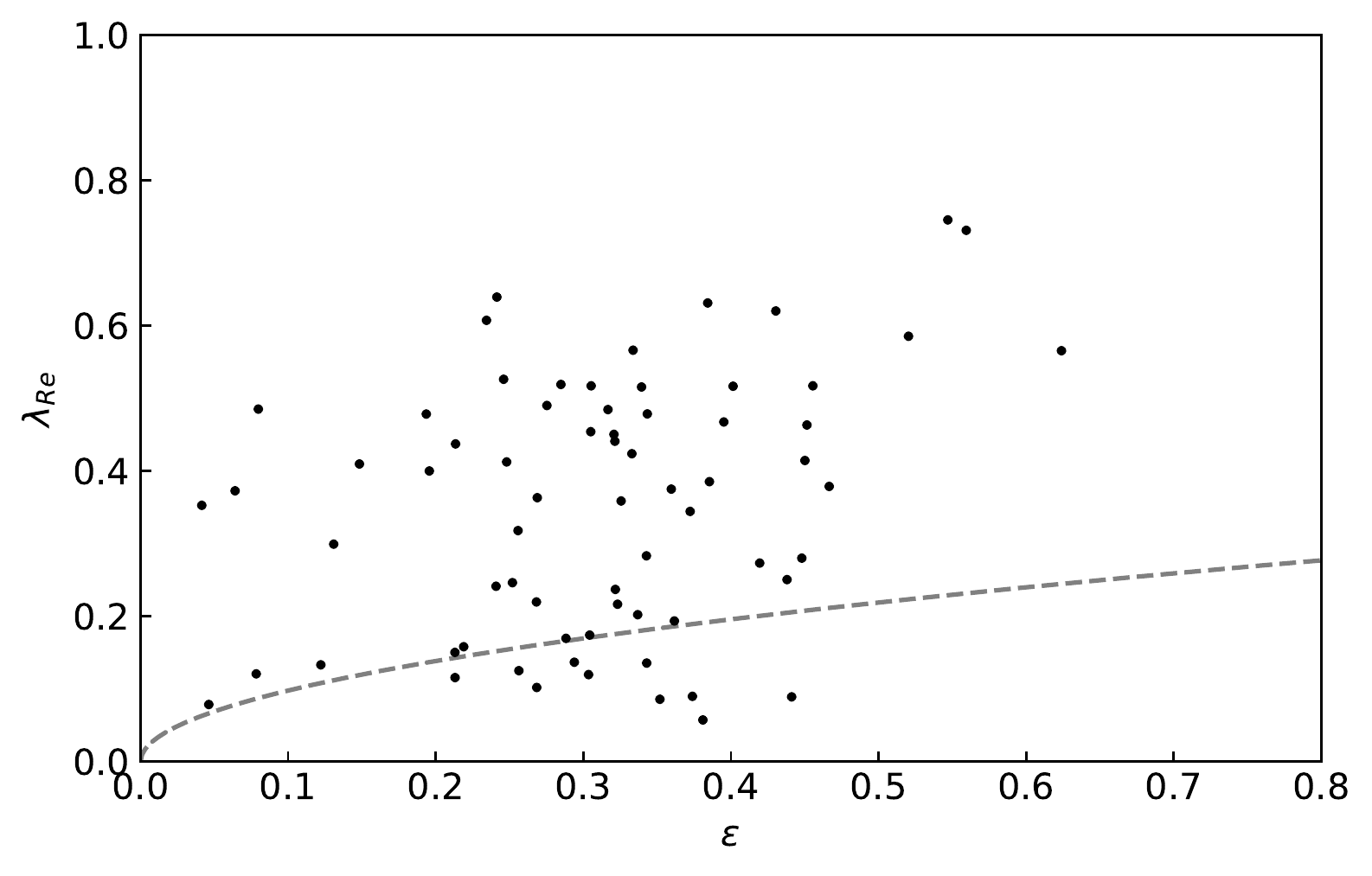}
 \caption{The proxy for the specific angular momentum within an aperture of $1~R_e$, $\lambda_{Re}$, versus
the ellipticity, $\epsilon$.  The dashed line represents the separation between slow and fast rotators (below and above the line, respectively) using the definition $0.31 \times \sqrt{\epsilon}$ from \citet{Emsellem_2011}.}
 \label{fig:kinematics}
\end{figure}

Another method that is commonly used to distinguish S0s from ellipticals is to use the kinematics to determine whether the galaxy is a fast or slow rotator, respectively. Following  \citet{Fraser_2021}, the spin parameter, $\lambda_{Re}$, which is a proxy for the specific angular momentum, was plotted against the ellipticity for each galaxy in Fig.~\ref{fig:kinematics}. The ellipticities were taken from the NSA, and were deprojected and PSF-corrected using the prescription of \citet{Harborne_2020}, while the $\lambda_{Re}$ values were calculated using the definition of  \citet{Emsellem_2011} within an elliptical aperture with a semi-major axis equal to $1~R_e$ (where this value was also taken from the NSA). The dashed line represents the separation between slow and fast rotators, defined as $0.31\times \sqrt{\epsilon}$ \citep{Emsellem_2011}. It can be seen that while the majority of the galaxies lie above this line in the fast-rotator regime, 10 galaxies lie below it in the slow rotator regime. An investigation of the properties of these galaxies (e.g. T-Type, P$_{\text{S0}}$, B/T light ratio, structural parameters etc) revealed no obvious offsets of these galaxies from the rest of the sample. The analysis and plots presented in the rest of this paper were also carried out without these slow rotators in the sample, with no significant difference in the results. Therefore, we chose to keep these galaxies in our sample and focus on the visual morphological classification.

The  MDLM-VAC gives 2 probabilities for the galaxies having bar signatures, one from Galaxy Zoo~2 \citep{Willett_2013} and the other from \citet{Nair_2010}, which uses data from the SDSS DR4 \citep{Stoughton_2002}. Using a limit of 0.5, Galaxy Zoo~2 indicates that 7 galaxies have a clear bar signature, while \citet{Nair_2010} only gives three galaxies with likely bars. Within these samples, only two galaxies appear in common. Therefore, the majority of the S0 galaxies used in this study can be considered unbarred.

\begin{figure}
 \includegraphics[width=\linewidth]{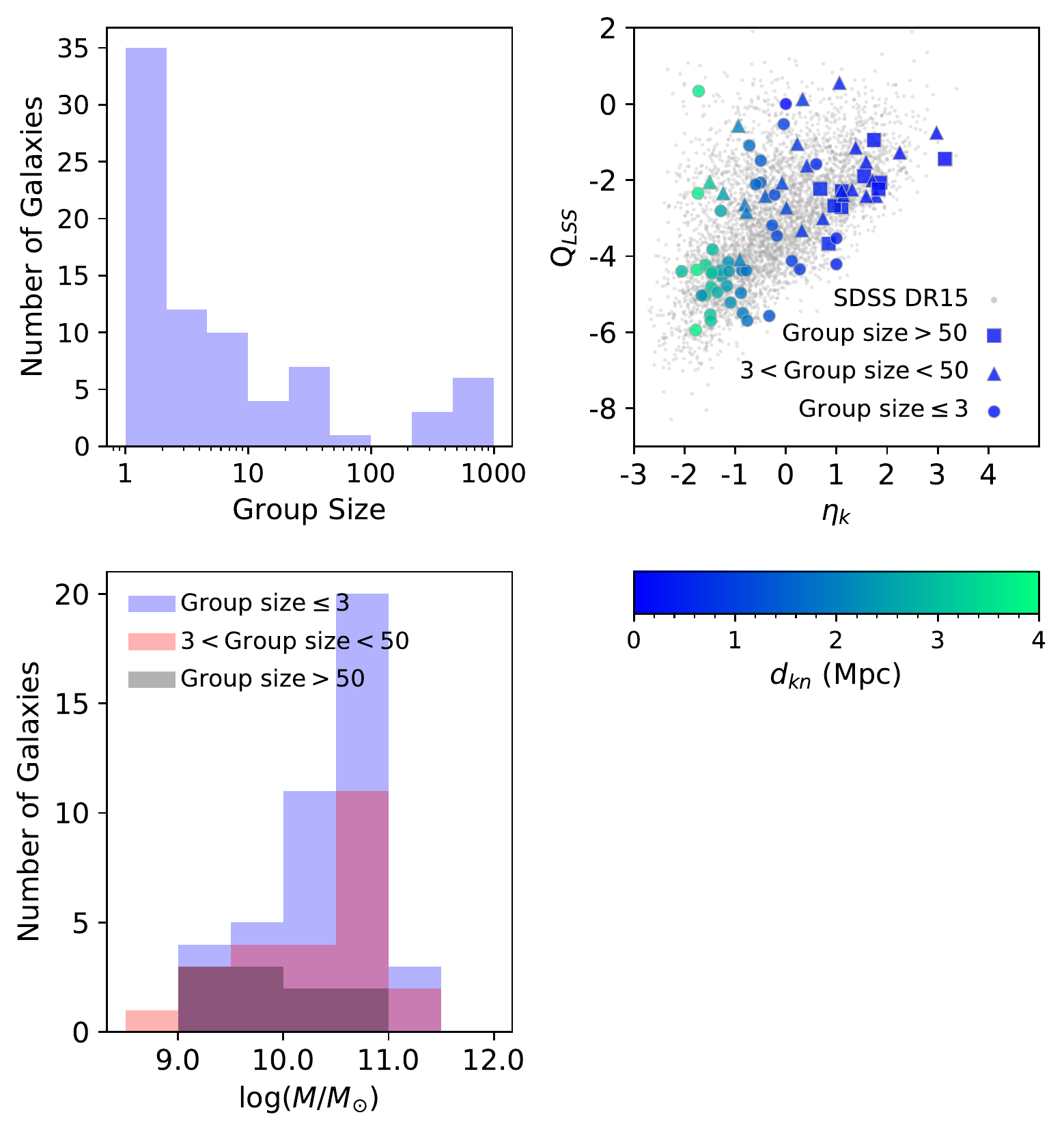}
 \caption{An overview of the environment distribution for the sample of galaxies. The histogram on the top-left shows the number of neighbouring galaxies in the group that each galaxy belongs to (group size = 1 corresponds to an isolated galaxy). The plot on the top-right shows the tidal strength parameter ($Q_{LSS}$) against the projected density ($\eta_{k,LSS}$). The colours represent the distance to the fifth nearest neighbour for each galaxy, and the grey points represent the measurements for all galaxies included in the GEMA-VAC.  The circles, triangles and squares represent galaxies identified as isolated and in groups and clusters respectively, and the mass distribution of the galaxies in these three environments are given in the histogram on the bottom-left. } \label{fig:sample_environment}
\end{figure}

The Galaxy Environment for MaNGA Value Added Catalog \citep[GEMA-VAC;][]{Argudo_2015} describes the environment for all galaxies in DR15, using CasJobs
tool to search for neighbours within 1~Mpc of each galaxy. Fig.~\ref{fig:sample_environment} shows the histogram for the group size distribution from this catalog, where the group size indicates the number of members of the group to which each galaxy belongs. Isolated galaxies have a group size of 1. It can be seen that the majority of galaxies belong to groups of $<10$~members, with $\sim45\%$ of the galaxies being isolated or in pairs. 

To further explore the environment for each galaxy, particularly those in larger groups, we look in more detail at the environment in terms of the large scale structure (LSS), also from the GEMA-VAC. These values consider all known neighbouring galaxies within a  volume of projected distance radius of 5~Mpc of each MaNGA galaxy and within a line-of-sight velocity difference of $\Delta$v~$\leq~500$~km~s$^{-1}$, as measured in either the SDSS main galaxy spectroscopic sample \citep{Strauss_2002} or the Baryon Oscillation Spectroscopic Survey \citep[BOSS,][]{Dawson_2013}. Three parameters have been selected as measures of the local environment- the tidal strength parameter ($Q_{LSS}$),  the projected density for the fifth nearest neighbour ($\eta_{k,LSS}$, where $k=5$), and the projected distance in Mpc to the fifth nearest neighbour ($d_{kn}$). The tidal strength represents an estimation of the total gravitational interaction strength that the neighbours produce on the central galaxy with respect to its internal binding forces, with lower values for $Q_{LSS}$ reflecting more isolation from the effects of neighbours, while the projected density characterises the LSS around the galaxy, with lower values for $\eta_{k,LSS}$ indicating larger distances to the $k$th nearest neighbour. Further details on these parameters can be found in \citet{Argudo_2015}. These three values are generally complementary of each other. For galaxies located in low density environments, one would expect $Q_{LSS}$ and $\eta_{k,LSS}$ to be small with larger $d_{kn}$, while galaxies in higher density environments would have larger $Q_{LSS}$ and $\eta_{k,LSS}$ and a smaller $d_{kn}$. A low $\eta_{k,LSS}$ and high $Q_{LSS}$ however indicates that the galaxy is a member of a small, isolated, compact group, while the opposite trend (i.e. high $\eta_{k,LSS}$ and low $Q_{LSS}$) could indicate that the galaxy belongs to a large group or small cluster with large distances between each member \citep{Lacerna_2018}. Fig.~\ref{fig:sample_environment} plots these environment measures for each galaxy in the sample as the coloured points. The sample has been split up into three environments- (relatively) isolated galaxies with 1, 2 or 3 members in their groups (circles, 43 galaxies), group galaxies with between 4 and 50 members (triangles, 25 galaxies), and cluster galaxies with more than 50 members (squares, 10 galaxies). The measurements for the full MaNGA DR15 sample plotted as grey points for comparison. One can see that the more isolated galaxies occupy regions of low  $\eta_{k,LSS}$ and $Q_{LSS}$ with larger $d_{kn}$, while those galaxies in larger groups and clusters lie towards  higher $\eta_{k,LSS}$ and  $Q_{LSS}$ with smaller $d_{kn}$, as expected. The mass distribution for the galaxies in each environment has also been plotted in Fig.~\ref{fig:sample_environment}, and show a slight bias such that the isolated and group galaxies tend to have higher masses than the cluster galaxies. However, it should be noted that the relatively low numbers in the group and, particularly, the cluster environments compared to the isolated galaxies, and the overlap in the tidal strength and projected density parameters, mean that any environmental trends seen in this work are tentative at best, and so these results should be used with caution.

Throughout this paper, the results presented were derived using the spectra from the SE fits. In general, the S\'ersic component represented the bulge while the exponential component was taken to reflect the disc. However, in around 5 galaxies in this sample, the components were found to be `flipped', where the S\'ersic and exponential components reflect the disc and bulge respectively. A full discussion of the flipped galaxies can be found in \paperI\., but in general a galaxy was assumed to be flipped if the $R_e$ of the exponential component was smaller than that of the S\'ersic component and the S\'ersic index was less than 1.

\section{Stellar Populations Analysis}\label{sec:stellar_pops}
The stellar populations analysis was carried out using two methods to give three sets of complementary results- luminosity-weighted populations derived through measurements of the line strengths using the Lick index definitions of \citet{Worthey_1994}, and luminosity and mass-weighted stellar populations derived through full spectral fitting with \textsc{ppxf}. The luminosity-weighted ages and metallicities provide information on the \textit{most recent episode} of star formation since this phase created the hottest, brightest stars present in the galaxy that dominate the light. In the case of quiescent galaxies, such as S0s, the luminosity-weighted ages indicate when the star formation was truncated, while the metallicities tell us about the chemical enrichment of the gas that fuelled that final episode. The mass-weighted stellar populations, on the other hand, give a better overview as to how the mass of the galaxy built up over time, and is less biased towards more recent episodes of star formation that dominate the light but may only account for a small fraction of the total mass of the galaxy. 

Together, the luminosity and mass-weighted stellar populations allow us to build up a clearer picture of the star-formation histories of the bulges and discs. It should be noted that this section focusses on describing the methods used and presenting a qualitative analysis of the results. A more quantitative analysis comparing the properties of the bulge and disc in each galaxy for both methods presented here will be presented in Section~\ref{sec:bulges_and_discs}.

\subsection{Stellar Populations from Line Strengths}\label{sec:line_strengths}
\citet{Fraser_2018b} carried out an analysis of the luminosity-weighted stellar populations of bulges and discs of S0s within the MaNGA Survey. In that study, they used photometric decomposition to determine the bulge and disc light fractions for all spaxels, and identified regions within the datacubes that were bulge and disc dominated to create the spectra representing those components. They selected their sample of S0s from the MaNGA product launch 5 (MPL-5, released as SDSS DR15), using kinematic selection techniques alongside traditional morphological classifications to select `smooth', fast-rotating galaxies. This technique resulted in a final sample of 279 S0 galaxies observed with all 5 IFU sizes.  Of these 279 galaxies, only 44 were observed with the 91 and 127-fibre IFUs used in this study. The final overlapping sample consisted of 26 galaxies that had good fits with \textsc{buddi}. Of the remaining 18, 6 galaxies had failed fits (i.e. \textsc{GalfitM} failed to converge on a good solution and crashed), and 12 galaxies were considered `poor' fits based on the structural parameters, as outlined in Section~\ref{sec:sub_decomp}.

Due to this overlap in the sample of \citet{Fraser_2018b} and the similarity in our goals, the first step in the analysis was to consider the luminosity-weighted stellar populations of the bulges and discs using the line strengths. While this analysis was carried out as a verification of the method and results, it also provides valuable information on the recent star-formation histories of the bulges and discs. 

\subsubsection{Luminosity-Weighted Ages and Metallicities}\label{sec:age_met}
The analysis of the stellar populations of bulge and disc-dominated regions of S0s in the MaNGA survey  carried out by \citet{Fraser_2018b} followed the procedure outlined in \citet{Parikh_2018} to correct for emission lines and velocity dispersion, and so in this work we have followed their methodology in order to better compare our results to that study. 

The first step was to model the spectra and subtract off any emission lines present. For this step, each spectrum was modelled with \textsc{ppxf} to fit both the stellar continuum and gas emission spectra. The stellar continuum was modelled using the MILES evolutionary synthesis models of \citet{Vazdekis_2015}.  The fits were also  convolved with multiplicative Legendre polynomial of order 10 to model for the shape of the continuum and make the fit insensitive to dust reddening. The emission lines were modelled as a series of Gaussian profiles at wavelengths of known emission features. Once a good fit was achieved, the peak of the H$\beta$ flux and the standard deviation in the residual spectrum, created by subtracting the best fit spectrum from the input spectrum, were measured. If the peak of the H$\beta$ flux was greater than 3 times the standard deviation in the residual spectrum, the best-fit emission line spectrum was subtracted off of the input spectrum to leave a pure stellar continuum spectrum. 

The emission-subtracted stellar continuum spectrum was then modelled again with \textsc{ppxf} to obtain two model spectra representing the stellar continuum, one at the resolution of the MILES models (which is similar to the SDSS resolution) and the other convolved with the velocity dispersion of the galaxy. The strengths of the H$\beta$, Mg$b$, Fe5270 and Fe5335 absorption lines were measured in all three spectra (gas-subtracted input spectrum, and convolved and unconvolved model spectra) with the \textsc{indexf} software of \citet{Cardiel_2010}. The uncertainties in these measurements derived through a series of simulations using the errors in the line-of-sight velocities and the S/N measured from the spectrum itself, details of which can be found in \citet{Cardiel_1998}. The ratios of the line strengths from the two model spectra were then measured, and used to correct line strengths measured from the gas-subtracted input spectrum to the SDSS resolution (i.e. correcting for the kinematic broadening of the lines). Finally the corrected line strengths were plotted onto the Simple Stellar Populations (SSP) model grids of \citet{Vazdekis_2010} in order to convert them into estimates of the luminosity-weighted stellar populations. The H$\beta$ index was used as an indicator of age, while $\text{[MgFe]$'$}=\sqrt{\text{Mg}b\ (0.72 \times \text{Fe}5270 + 0.28 \times \text{Fe}5335)}$ was used as a metallicity indicator. The model grid was created using the  MILES web-based tool\footnote{http://miles.iac.es/}, which measures the line strengths from the MILES stellar library \citep{Sanchez_2006} for single stellar populations covering a wide range in age ($0.50-15.85$~Gyr) and metallicity ($-2.32$ < [M/H] < $0.22$). To account for the small difference in the instrumental spectral resolution between the MILES and MaNGA spectra, the tool convolved the spectra in the stellar library with a gaussian of the appropriate dispersion before creating the final SSP model grid, however this step made very little difference in the results.

\begin{figure}
 \includegraphics[width=\linewidth]{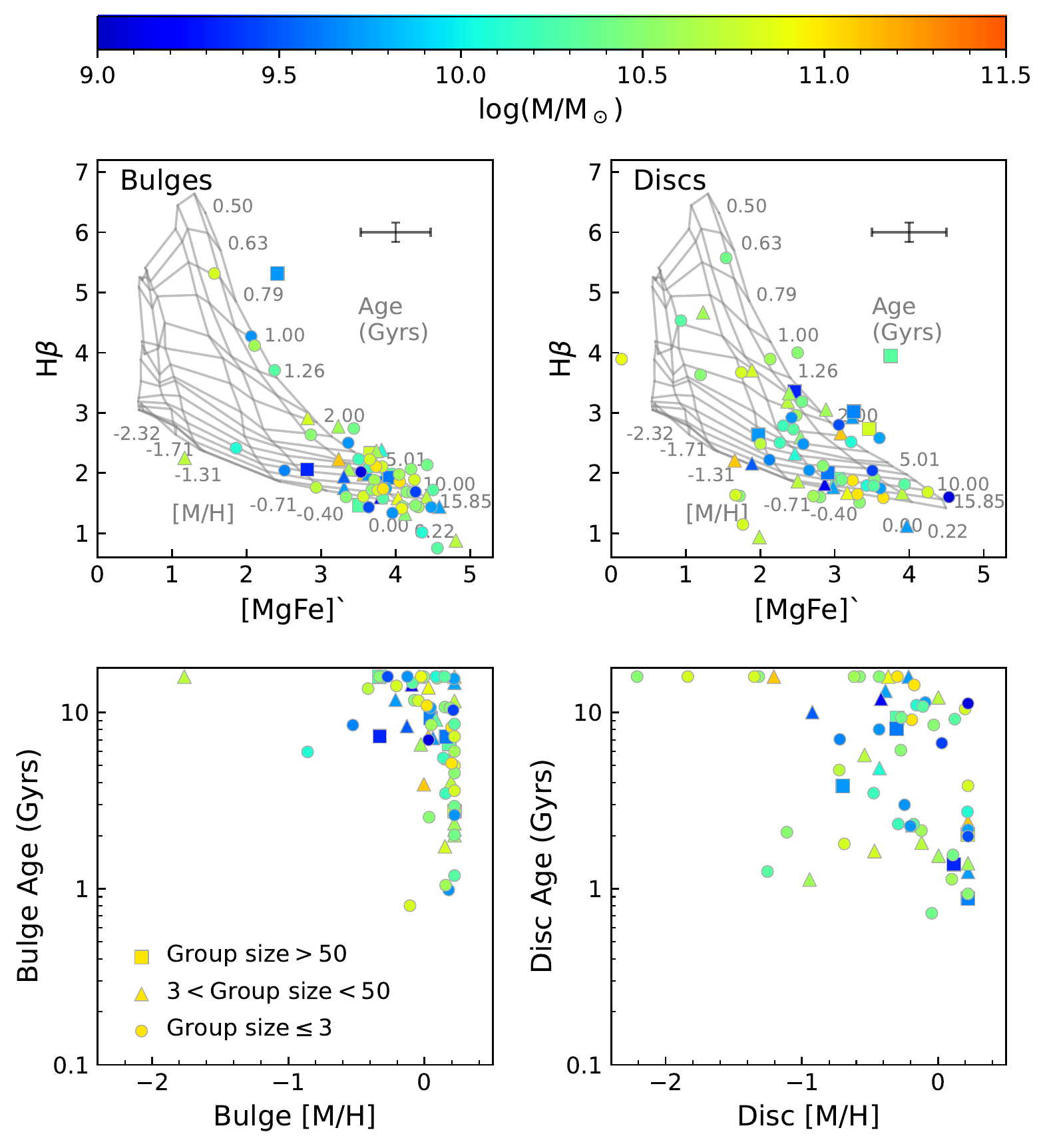}
 \caption{Light-weighted stellar populations of the bulges and discs derived using the Lick indices. The line strengths measured from the bulge and disc spectra are plotted onto the SSP model grids in the top row, and the extrapolated ages and metallicities are plotted below for the bulges (left) and discs (right). In all plots, the colours reflect the total mass of the galaxy, as shown by the colour bar, and the shapes represent the environment according to the legend.
} \label{fig:stellar_pops}
\end{figure}

Figure~\ref{fig:stellar_pops} shows the line strength measurements for the bulges and discs plotted onto the SSP models, with the ages and metallicities derived through interpolation for each component below. The points have been colour-coded according to the total mass of the galaxy. It can be seen that the bulges tend to lie in the old, metal-rich part of the SSP model grid, with the bulges of more massive galaxies showing higher values for their ages. The bulge age and metallicity plot below also reflects this trend, where the majority of bulges with younger and more metal poor populations also belong to lower mass galaxies.  The discs, on the other hand, show more of an extension towards the younger, more metal-poor regime, but display little or no trends with galaxy mass. 

\begin{figure}
 \includegraphics[width=\linewidth]{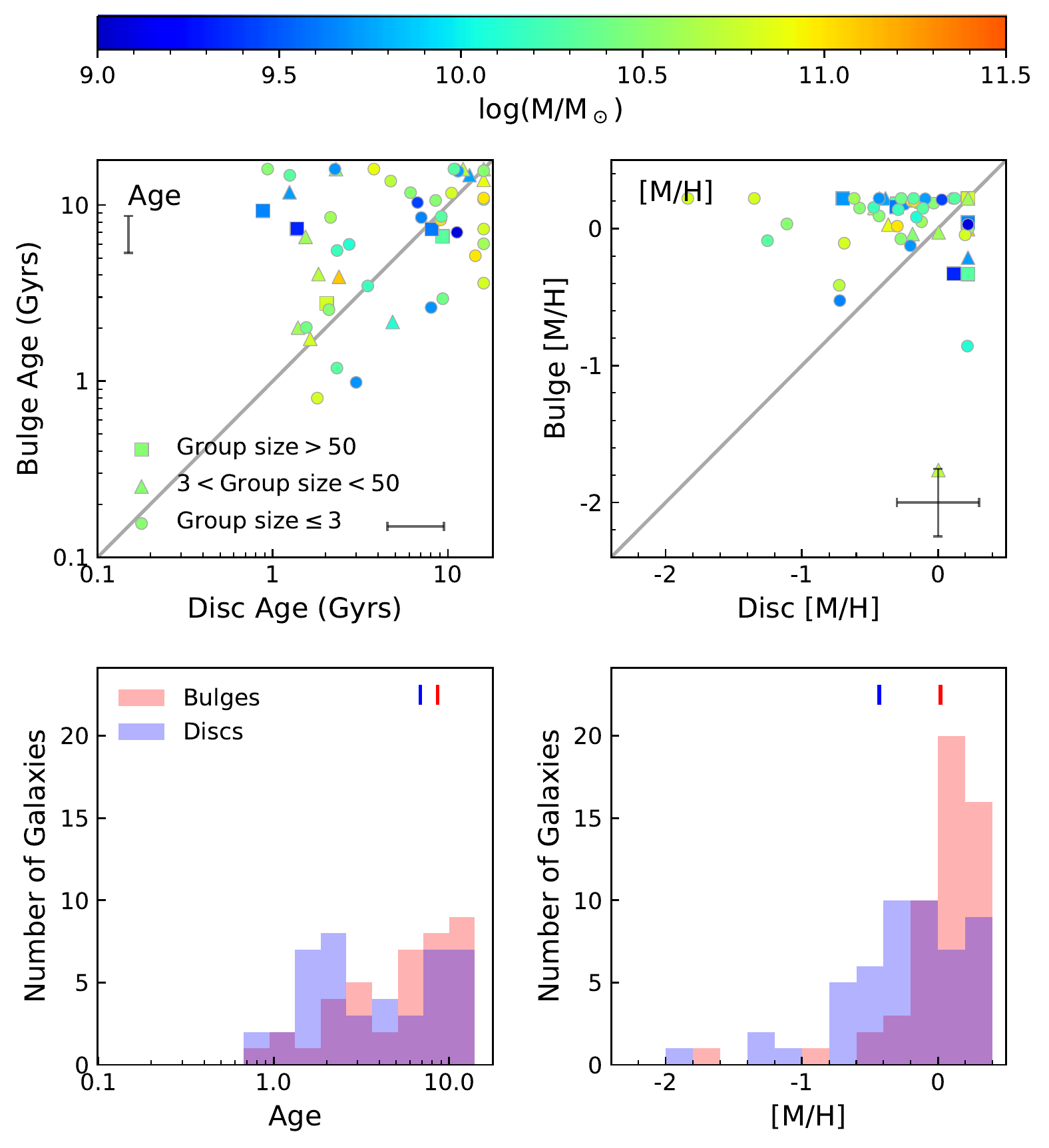}
 \caption{\textit{Top}: A comparison of the bulge versus disc luminosity-weighted stellar populations derived from the Lick indices, with the ages on the left and the metallicities on the right. The mean uncertainties have been plotted at an age of 7~Gyrs for each axis in the age plot, and in the bottom right for the metallicity plot. The triangles and circles represent galaxies in groups of $>3$ or $\leq3$ respectively. \textit{Bottom}: Histograms showing the distribution of the bulge and disc ages and metallicities. 
} \label{fig:stellar_pops_BvD}
\end{figure}

The top row of Fig.~\ref{fig:stellar_pops_BvD} presents a comparison of the ages and metallicities in the bulges and discs of each galaxy, and shows that bulges generally display older ages and higher metallicities than their corresponding discs. One can see that while some bulges do show evidence of younger or more metal-poor stellar populations than their corresponding discs, generally these measurements are still consistent within their uncertainties. This figure also presents the distributions of the bulge and disc ages and metallicities as histograms, where the mean values for their ages and metallicities are marked with bars at the top of the plots and listed in Table~\ref{tab:ages_mets}. A potential bimodality can be seen in the disc ages, though with the small sample used here it's unclear whether this effect is real or a coincidence.

\begin{table}
	\centering
	\caption{The mean mass-weighted (MW) and luminosity-weighted (LW) ages and metallicities and the associated standard deviation for the  spectra extracted for the bulges and discs in the SE fits. In the method column, \textsc{ppxf} refers to measurements derived using full spectral fitting with \textsc{ppxf} and Lick corresponds to measurements of the line strengths and extrapolation from the SSP models.}
	\label{tab:ages_mets}
	\begin{tabular}{lrrrrr} 
		\hline
		  &  Method & Bulges & Discs   \\
		\hline
LW Age (Gyrs)		& Lick 		& $9\pm 5$	& $7\pm 5$ 	\\
LW Age (Gyrs)		& \textsc{ppxf} 	& $8\pm 4$	& $5\pm 3$ 	\\
MW Age (Gyrs)		& \textsc{ppxf} 	& $10\pm 3$	& $7\pm 3$ 	\\
\\
LW [M/H]			& Lick	 	& $0.0\pm 0.3$		& $-0.4\pm 1.0$ 	\\  
LW [M/H]			& \textsc{ppxf} 	& $-0.03\pm 0.40$	& $-0.4\pm 0.4$ 	\\
MW [M/H]			& \textsc{ppxf} 	& $0.06\pm 0.40$	& $-0.3\pm 0.4$ 	\\

		\hline
	\end{tabular}
\end{table}

As with any studies of unresolved stellar populations, age-metallicity degeneracy is still an open problem \citep[See review by][and references therein]{Conroy_2013}. While the use of line strength indices helps break this degeneracy somewhat when compared to photometric stellar populations analysis, particularly for spectra with S/N$>100$\AA\ \citep{Trager_2000, Kuntschner_2000}, it is still not perfect for lower S/N spectra. For example, \citet{Worthey_1994} showed that the spectrum from an unresolved stellar population can be almost indistinguishable from that of another unresolved population with three times the age and half the metallicity. Additionally, simulations of mock galaxy spectra by \citet{Thomas_2005} and \citet{Sanchez_2011} have revealed an anticorrelation between age and metallicity measurements from the Lick indices, such that galaxies with older ages were also found to contain lower metallicities. While no perfect solution to this problem exists, in general the age-metallicity can be broken to some extent by using a combination of the hydrogen Balmer lines in conjunction with abundance rations related to Iron, such as [Mg/Fe]$^\prime$, or through full spectral fitting \citep[See Section~\ref{sec:full_spec_fitting};][]{Sanchez_2011}.

The SSP model grid in Fig.~\ref{fig:stellar_pops} plots the H$\beta$ versus [Mg/Fe]$^\prime$ indices, and demonstrates that the differences in the H$\beta$ line strengths decreases with increasing age for a constant metallicity, making stellar populations with ages $>10$~Gyrs difficult to distinguish. As a result, this issue may have an impact on the bulge and disc stellar populations, particularly when comparing the ages and metallicities of components in the same galaxy, such as in Fig.~\ref{fig:stellar_pops_BvD}.  The mean uncertainties on the line strength measurements are given above the SSP model grid, and it can be seen that particularly in the older stellar populations, the error bars cover multiple lines of constant age and metallicity in the SSP model grids. Furthermore, the corresponding mean uncertainties in age and metallicity are given in the top panels of Fig.~\ref{fig:stellar_pops_BvD}. Therefore, in an effort to reduce the effect of the age-metallicity degeneracy in our results and conclusions, we present a more quantitative analysis of these measurements in Section~\ref{sec:bulges_and_discs}. That section explores the cases where bulges are significantly older/younger and more metal-rich/poor than their surrounding discs, where they are considered to be truly and significantly different if the age and metallicity measurements are inconsistent according to the uncertainties.

\subsubsection{Luminosity-Weighted $\alpha$-enhancement}\label{sec:alpha}
To learn more about the most recent episode of star formation in the bulges and discs, we can use the $\alpha$-element enhancement  to derive information on the star-formation timescale of that event and the origin of the gas that fuelled it.  The $\alpha$-enhancement is typically measured as the ratio of an $\alpha$-element-sensitive indices, in this case Mg$b$ to $\langle \text{Fe} \rangle$, where $\langle \text{Fe} \rangle=(\text{Fe}_{5270}+\text{Fe}_{5335})/2$. Type~II supernovae are the main source of $\alpha$-element enrichment of the interstellar medium, while Fe originates mainly in Type~Ia supernovae. Since Type~II supernovae typically start exploding shortly after the onset of star formation while Type~Ia supernovae only start $\sim1$~Gyr later, the ratio of $\alpha$ elements to Fe can be used as an indicator of the star-formation timescale. For example, shorter episodes of star formation will give an $\alpha$-enhanced stellar population due to the enrichment of magnesium from the SNII, and the $\alpha$-enhancement will begin to drop after SNIa appear due to the dilution of magnesium with iron in the interstellar medium. It has been found that the highest [$\alpha$/Fe] ratios are achieved in galaxies with the shortest  half-mass formation time \citep[<2~Gyr;][]{delaRosa_2011}, and that the [$\alpha$/Fe] decreases with increasing star-formation timescale out to $\sim10\,$Gyr \citep{Wiersma_2009}.

\begin{figure*}
 \includegraphics[width=\linewidth]{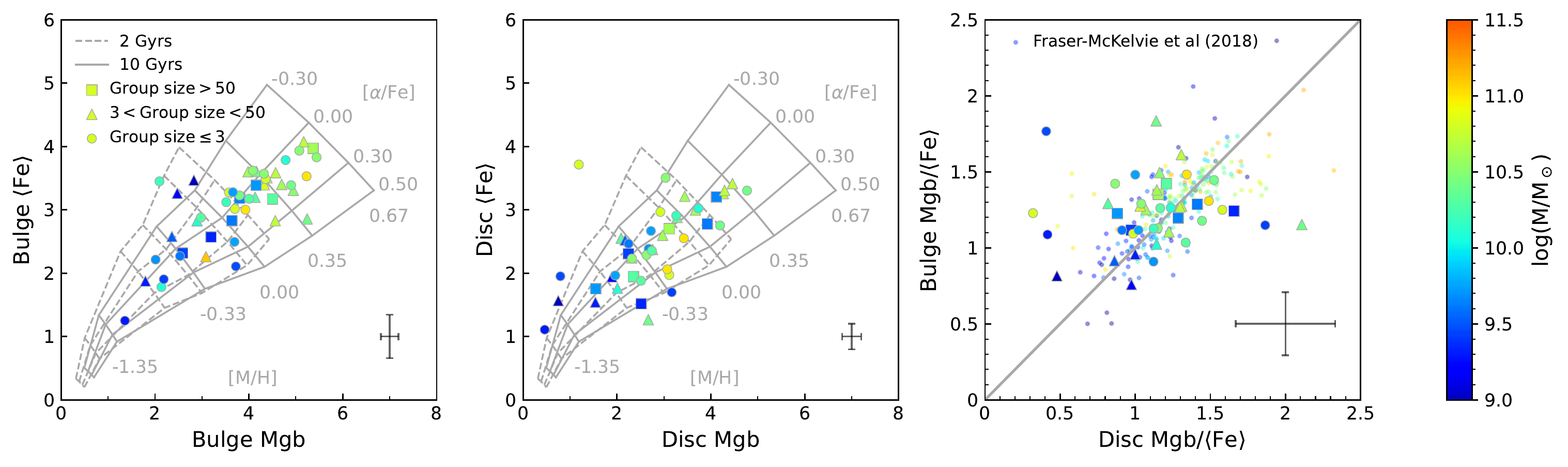}
 \caption{Mgb and $\langle$Fe$\rangle$ line strengths for the bulge (left) and disc (middle), with the models of \citet{Thomas_2011}  for 2 and 10~Gyrs overplotted as dashed and solid lines repsectively for reference. On the right is a comparison of the Mgb/$\langle$Fe$\rangle$ ratios for the bulges and discs, showing that in general they are relatively consistent. The measurements from \citet{Fraser_2018b} have been plotted as smaller dots for comparison. In all plots, the error bar on the bottom-right represents the median uncertainty on the data points, and the colours represent the stellar mass, as shown in the colour bar on the right. 
 } \label{fig:stellar_pops_alpha}
\end{figure*}

In Fig.~\ref{fig:stellar_pops_alpha} the Mg$b$ and $\langle \text{Fe} \rangle$ line strengths are plotted onto the SSP model grids of \citet[][hereafter TMJ models]{Thomas_2011}, which give estimates of the relative metallicity and [$\alpha$/Fe] ratio for stellar populations of age 2 and 10~Gyrs. Since the \citetalias{Thomas_2011} models are based on the MILES stellar library, which has a similar spectral resolution to the MaNGA data, we followed \citet{Thomas_2011} and applied no corrections to the spectra to correct for the instrumental resolution, and only applied the kinematic corrections  discussed in Section~\ref{sec:line_strengths}. It can be seen that the bulges and discs show similar trends in their $\alpha$-enhancement with total stellar mass, such that higher mass galaxies show higher values for the Mg$b$ and $\langle \text{Fe} \rangle$ line strengths in both their bulge and disc, and thus higher metallicities. A comparison of the Mg$b/\langle \text{Fe} \rangle$ ratios for the bulges and discs is also given in the third panel of Fig.~\ref{fig:stellar_pops_alpha}, and shows that the $\alpha$-enhancement of the bulges and discs in each galaxy are consistent, regardless of the stellar mass. A similar result was seen in  \citet{Fraser_2018b}, which is also shown in Fig.~\ref{fig:stellar_pops_alpha} as the smaller data points, and in \citet{Johnston_2014}.

\subsection{Stellar Populations from Full Spectral Fitting}\label{sec:full_spec_fitting}
\subsubsection{Mass- and Luminosity-Weighted Ages and Metallicities}\label{sec:age_met_fsp}

While the line strengths give us a good indicator of the properties of the most recent episodes of star formation in the bulges and discs, full spectral fitting can provide more details on their star-formation history and mass assembly. Additionally, this method is also able to break the age-metallicity degeneracy better due to the increased information available for the fits \citep{Macarthur_2009, Sanchez_2011}. We used \textsc{ppxf} to derive estimates of the mass and luminosity-weighted stellar populations of the bulges and discs through this technique. This process uses a series of template spectra of known relative ages and metallicities, and finds the linear combination that best fits the galaxy spectrum. We again used the MILES evolutionary synthesis models of \citet{Vazdekis_2015}, created using the BaSTI isochrones \citep{Pietrinferni_2004,Pietrinferni_2006}, as the model spectra in this study. The fits were  convolved with a multiplicative Legendre polynomial of order 10 to account for the shape of the continuum, reducing the sensitivity to dust reddening, and omitting the requirement of a reddening curve \citep{Cappellari_2017}. Additionally, the fits were regularized, which acted to smooth the variation in the weights of templates with similar ages and metallicities. Furthermore, the process of regularization is able to further reduce the effect of the degeneracy between age and metallicity.  \citet{Boecker_2020} describes how full spectral fitting alone can be both ill posed (i.e. the combination of template spectra used for a solution may not be unique) and ill conditioned (fluctuations in the S/N can significantly affect the templates selected, and thus the measurement of age and metallicity), but the smoothing due to regularization is able to treat both issues.

The degree of smoothing was determined for each spectrum independently. The first step was to run an unregularized fit, after which the noise was scaled appropriately until $\chi^2/N_{DOF} = 1$, where $N_{DOF}$ is the number of degrees of freedom in the fit (i.e. the number of unmasked pixels in the input spectrum). The fit to the spectrum was then repeated using the scaled noise spectrum and increasing the regularization value until the $\chi^2$ of the fit increased by $\Delta\chi^2 = \sqrt{2\times N_{DOF}}$. This increase reflects the point at which the smoothed fit  still reflects the star-formation history of the galaxy, and before it has been smoothed too much. However, it is important to note that in reality, star-formation activity often varies over shorter timescales than the age steps of the models. Consequently, the results from the regularized fits are designed to reduce the age-metallicity degeneracy between spectra, thus allowing a more consistent comparison of systematic trends in their star-formation histories, and may not reflect the true star-formation history.

Estimates of  the mean luminosity and mass-weighted stellar populations for each spectrum were calculated from the weights of the template spectra using 
\begin{equation} 
	\text{log(Age)}=\frac{\sum \omega_{i} \text{log(Age$_{\text{template},i}$)}}{\sum \omega_{i}}
	\label{eq:age}
\end{equation}
and 
\begin{equation} 
	\text{[M/H]}=\frac{\sum \omega_{i} \text{[M/H]}_{\text{template},i}}{\sum \omega_{i},}
	\label{eq:met}
\end{equation}
respectively, where $\omega_{i}$ represents the weight of the $i^{th}$ template (i.e. the value by which the $i^{th}$  stellar template is multiplied to best fit the galaxy spectrum), and [M/H]$_{\text{template},i}$ and Age$_{\text{template},i}$ are the metallicity and age of the $i^{th}$ template respectively. The uncertainties on these measurements were calculated by taking the best fit to the stellar continuum for each spectrum, adding a random noise to result in the same Signal-to-noise ratio as the original spectrum, and modelling this new spectrum again with \textsc{ppxf} in the same way as before. This processes was repeated 50 times for each input spectrum, and the standard deviation in the measured ages and metallicities was taken as the uncertainty on the measurements from the bulge or disc spectrum.

\begin{figure}
 \includegraphics[width=\linewidth]{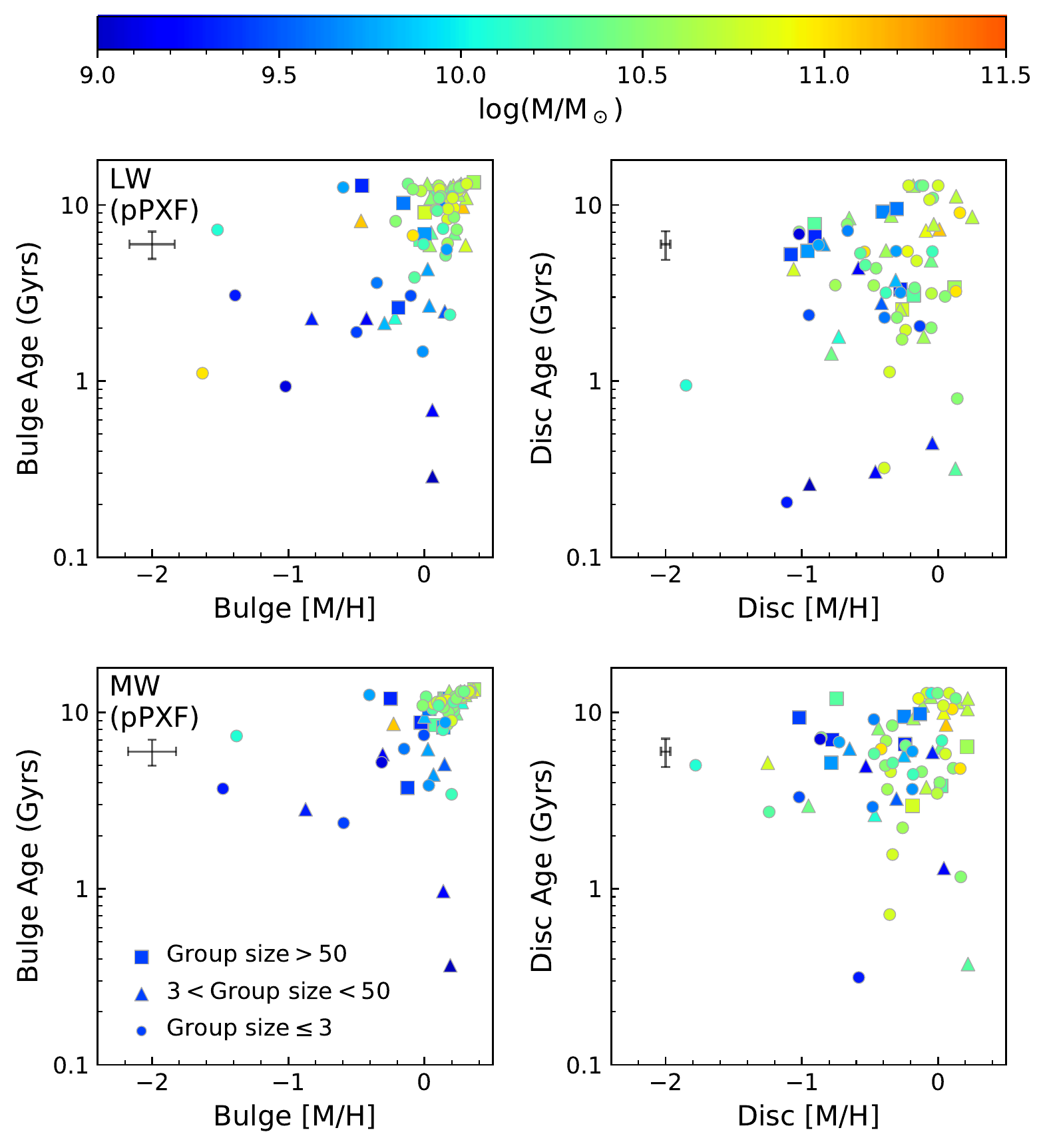}
 \caption{An overview of the ages and metallicities derived for the bulges (left) and discs (right) using \textsc{ppxf}. The top row shows the luminosity-weighted (LW) stellar populations, while the bottom row shows the mass-weighted (MW) populations. The colours reflect the total mass of the galaxy, as given by the colour bar, while the shapes represent the environment according to the legend.  The mean uncertainty in the measurements is given by the error bars on the left of each plot. 
 } \label{fig:stellar_pops_LWMW}
\end{figure}

The stellar populations derived through this technique are presented in Fig.~\ref{fig:stellar_pops_LWMW}, showing the luminosity-weighted ages and metallicities in the top row and the mass-weighted properties in the bottom row for bulges and discs on the left and right respectively. The mean uncertainty for each plot is given in the upper left of the plots. Again, the data points are coloured according to the total mass of the galaxy. For both analyses, a clear trend can be seen in the bulge properties with the mass of the galaxy, such that bulges in the higher mass galaxies tend to have old, metal-rich stellar populations while those in lower-mass galaxies extend towards younger ages and lower metallicities. The discs show a tentative trend with metallicity for both analyses, where lower-mass galaxies have lower metallicities. Interestingly, a tight correlation can be seen in the mass-weighted ages and metallicities of the bulges of galaxies with masses $>10^{10}M_\odot$, however given the difficulty in determining the ages of stellar populations above 10~Gyrs, it's uncertain how real this trend is. The old mass-weighted ages of these bulges indicate that they formed the majority of their mass a long time ago, early in their lifetimes, and this  tight correlation between age and metallicity may indicate that they formed over shorter timescales with few episodes of significant star formation.

\begin{figure}
 \includegraphics[width=\linewidth]{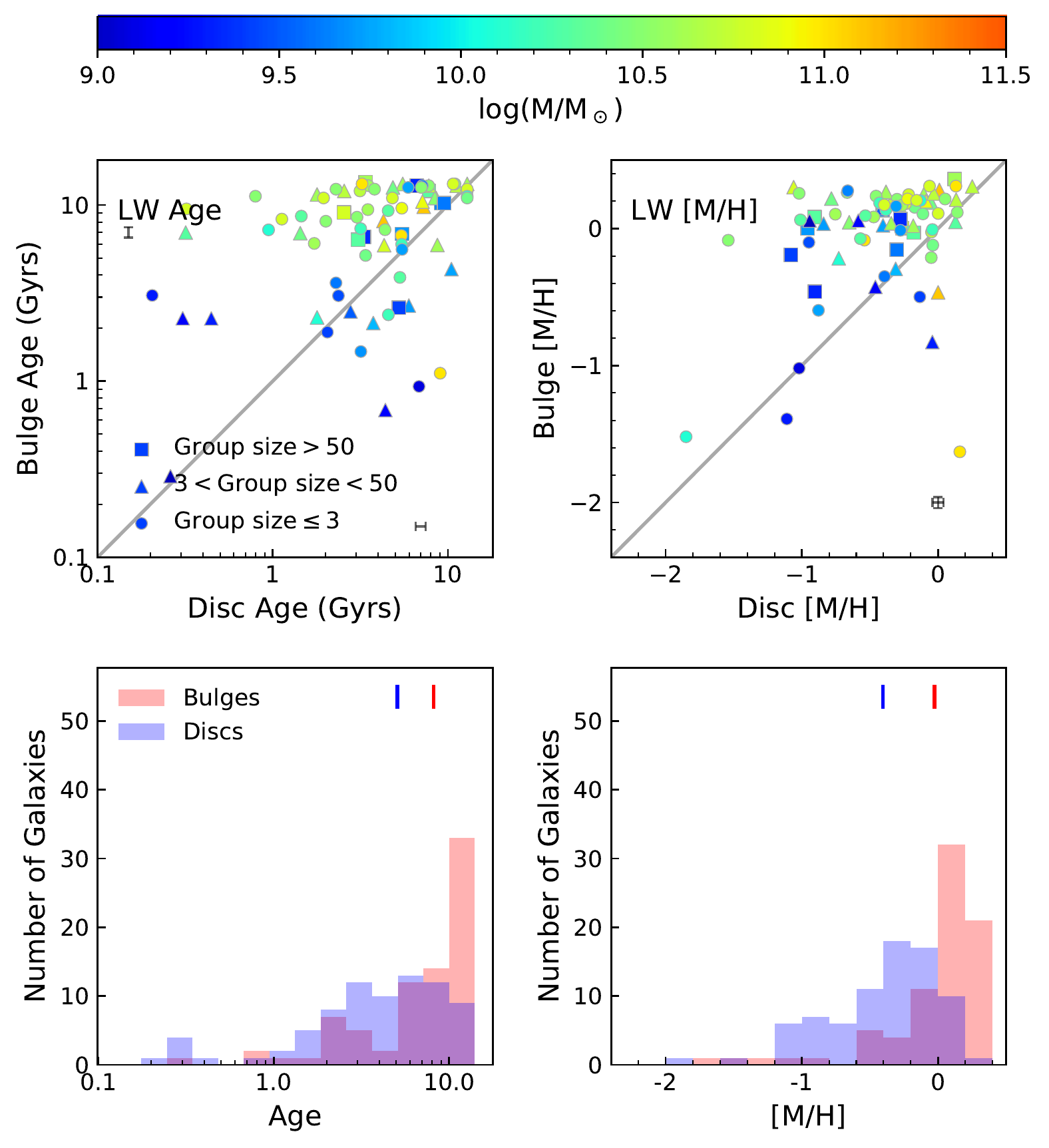}
 \caption{A comparison of the luminosity-weighted ages (left) and metallicities  (right) of the bulges and discs in each galaxy (top).  The mean uncertainty in the measurements is given by the error bars in  each plot. The histograms in the bottom row show the distributions in the ages and metallicites of the bulges (red) and discs (blue), with the vertical lines above the histograms showing the mean values for the bulges and discs.
} \label{fig:stellar_pops_LW}
\end{figure}

\begin{figure}
 \includegraphics[width=\linewidth]{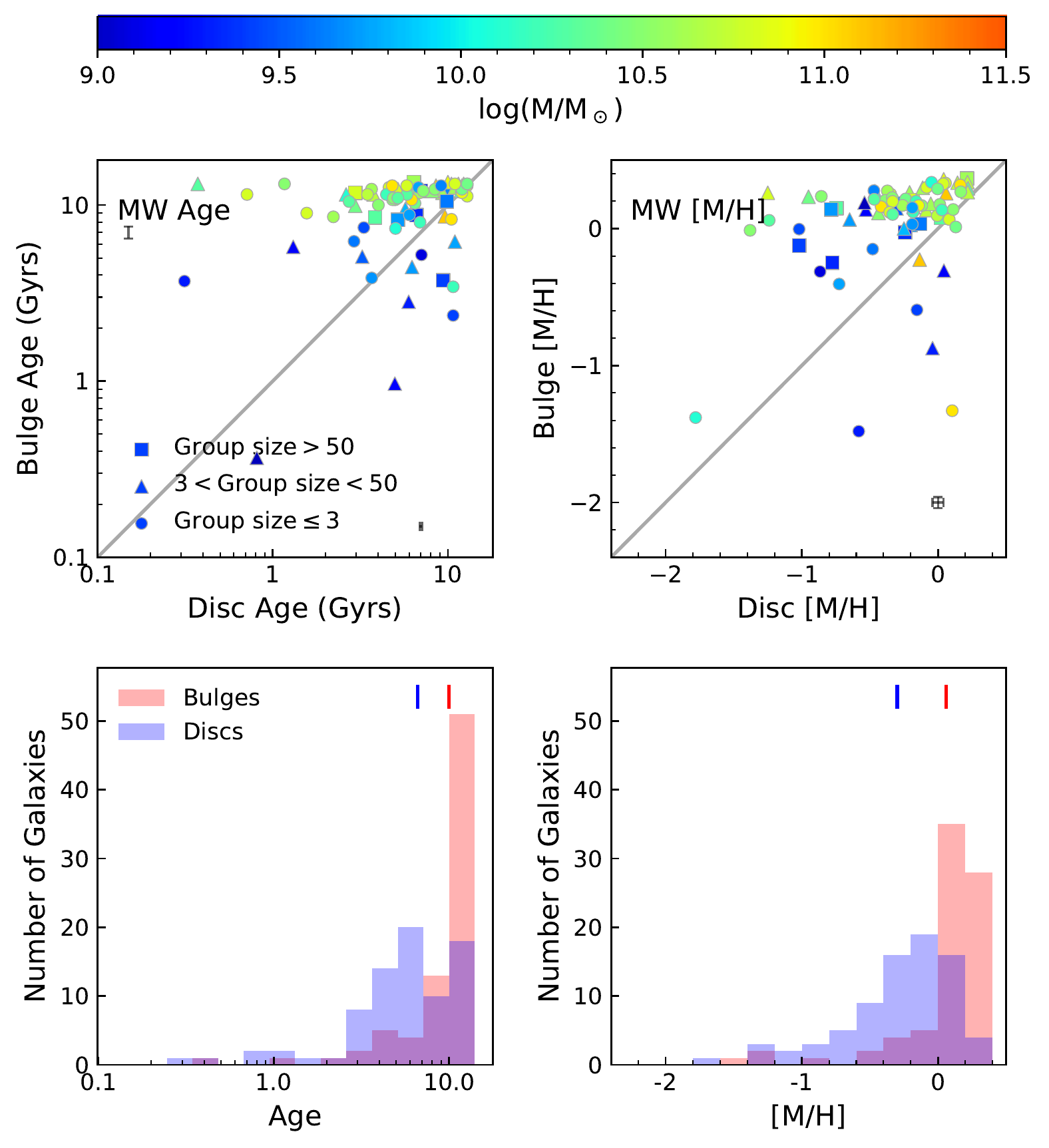}
 \caption{As Figure~\ref{fig:stellar_pops_LW}, but for the mass-weighted stellar populations.   
} \label{fig:stellar_pops_MW}
\end{figure}

When comparing the mass and luminosity-weighted properties of the bulges and discs, both components show evidence of younger and more metal poor luminosity-weighted populations for all masses. This trend can also be seen in Figures~\ref{fig:stellar_pops_LW} and \ref{fig:stellar_pops_MW}, which compare the luminosity and mass-weighted populations in the bulges versus the discs, respectively. The ages (left) and metallicities (right) of the bulges and discs are given in the top panels, and the histograms in the bottom panels give the distribution in these properties. As in Fig.~\ref{fig:stellar_pops}, the bars above the histograms represent the mean values for age and metallicity for the bulges and discs, and which are listed in Table~\ref{tab:ages_mets} along with the associated standard deviation. Together, these plots demonstrate that, in most cases, the stars that dominate the light in both components are not the same stars that dominate the mass, thus indicating that both components underwent a relatively extended star formation history and did not form completely in a single event.

\subsubsection{Star-Formation Histories}\label{sec:SFHs}
\begin{figure}
 \includegraphics[width=\linewidth]{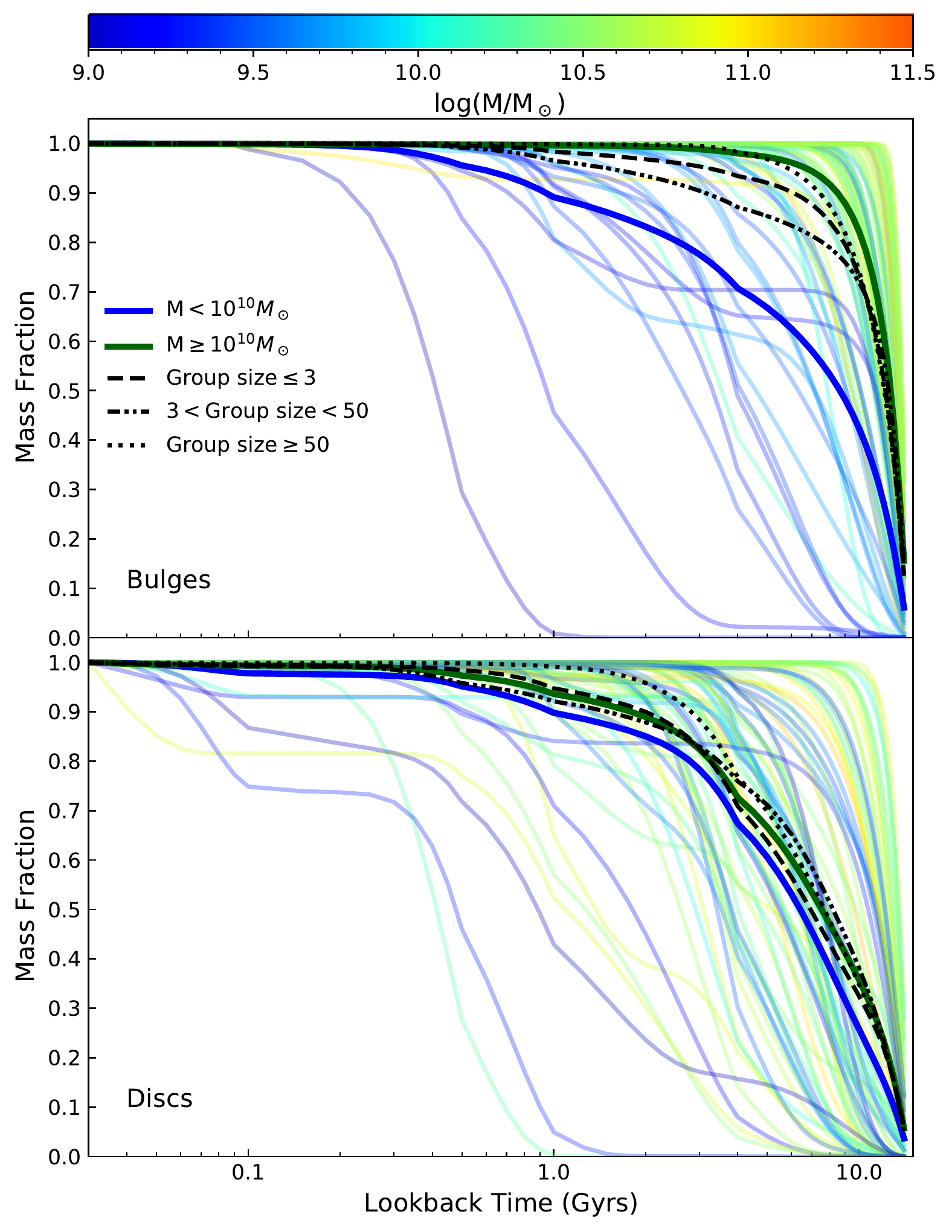}
 \caption{The mass assembly of the bulges (top) and discs (bottom) as a function of lookback time. In both plots, the thick blue and green lines show the mean mass-assembly profile for each component in galaxies with masses of $<10^{10}M_\odot$ and $>10^{10}M_\odot$ respectively, while the black dashed, dot-dash and dotted lines reflect the mean mass-assembly histories for each environment, according to the legend.
} \label{fig:SFHs}
\end{figure}

As well as giving the mean mass- and luminosity-weighted ages and metallicities, the weights of the stellar templates can also provide an estimate of the star-formation histories of the bulges and discs, thus giving an idea of how the mass in each component was assembled over the lifetime of the galaxy. Fig.~\ref{fig:SFHs} shows the mass-assembly history for all bulges and discs, giving the cumulative mass growth as a function of lookback time. These plots were created using the mass-weighted regularized fits by plotting the sum of the weights of all the template spectra used in each fit at each age step and older, and thus give an idea of how the mass within the galaxy was assembled over time. The colours again represent the total stellar mass of the galaxy, and the thick blue and green lines represent the mean mass-assembly histories of each component in galaxies with masses $<10^{10}M_\odot$ and $\geq 10^{10}M_\odot$, respectively.
These plots show that the discs have more extended SFHs than the bulges, as expected. However, the bulges show an interesting trend with galaxy mass, where bulges in higher mass galaxies assembled the majority of their mass over shorter timescales long ago while those in lower mass galaxies formed over more extended timescales and more recently.  No such trend is seen with the galaxy mass in the discs, where the mean mass-assembly histories of the discs of high and low mass galaxies are very similar. 

\begin{figure}
\begin{center}
\includegraphics[angle=0,width=1\linewidth]{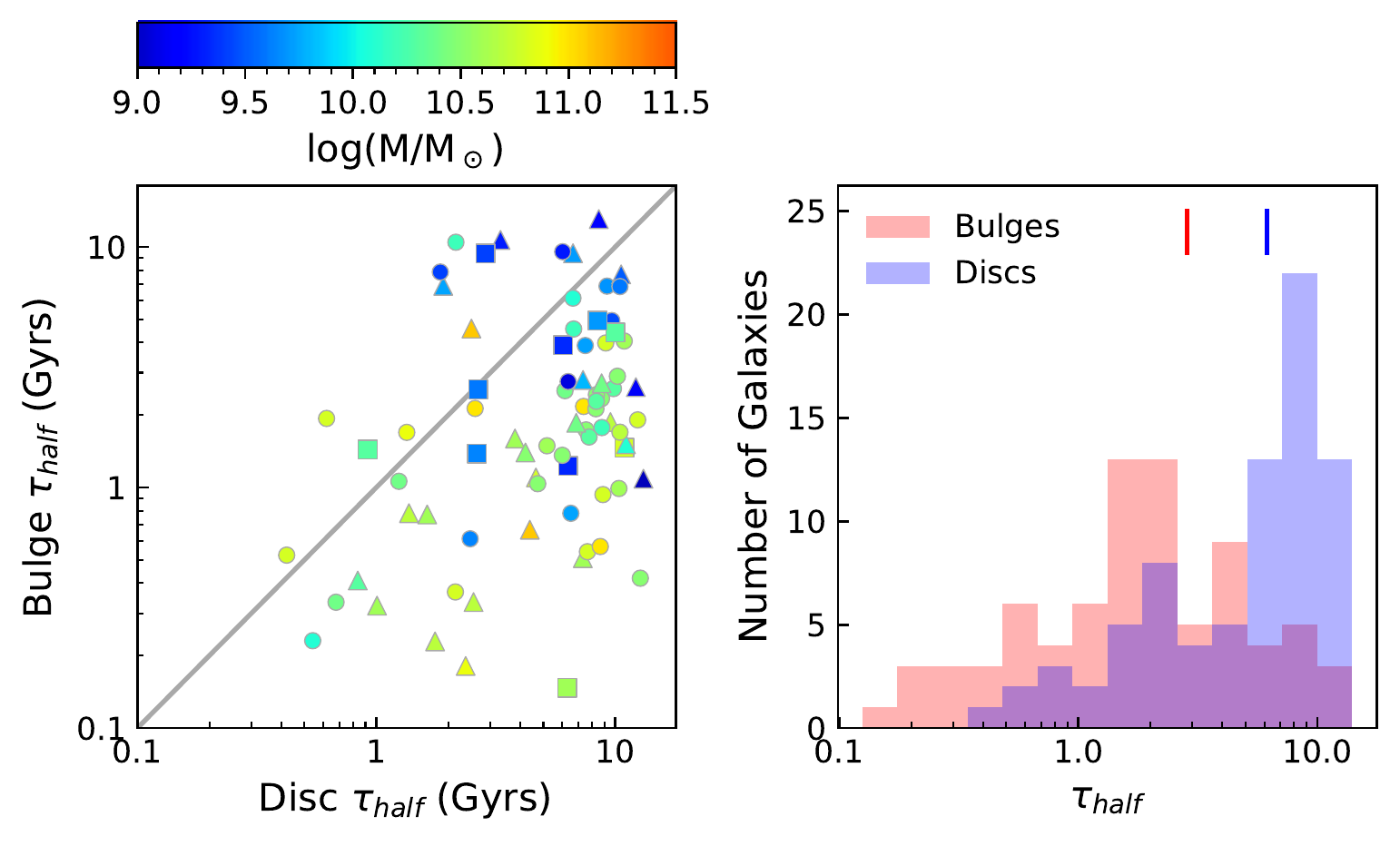}
\caption{Left: The formation timescales for the bulges and discs, plotted as the time taken to form half the mass of each component, $\tau_{half}$, colour coded according to the total mass of the galaxy according to the colour bar, and with squares, triangles and circles representing the environment (Group size $>$ 50, between 3 and 50, and $<$ 50 respectively, as in previous figures). Right: The histogram showing the distribution in the formation timescales of each component.
 \label{fig:formation_times}}
\end{center}
\end{figure}

Another way to display this information is to plot the formation timescales of each component. Figure~\ref{fig:formation_times} presents the formation timescales of the bulges and discs, $\tau_{half}$, which have been taken as the time taken to form half the mass of that component (i.e. by identifying the time difference between the oldest stellar template used in the fit with a non-zero weight and the lookback-time at which $50\%$ of the mass of the galaxy had assembled). The histogram clearly shows that the bulges experienced a shorter formation timescale than the discs, as discussed above. Similarly, the scatter plot also shows the correlation with total mass of the galaxy in the bulges, where bulges in higher mass galaxies formed over shorter timescales than those in lower mass galaxies. the discs however show little correlation with mass, except a tentative trend that all discs with $\tau_{half}\lesssim2$~Gyrs have masses $> 10^{10}M_\odot$.

In general however, the mass-assembly histories for the individual bulges and discs generally show a wide variety of trends. This result reflects a great diversity in the star-formation histories of these bulges and discs, indicating that their evolution is not completely dependent on the galaxy mass or environment, but also on the particular interactions of each galaxy during its life. For example, if real, the jumps seen in the cumulative mass assemblies of some bulges and discs over short times may indicate interactions, such as minor mergers or harassment. The accretion of a satellite galaxy in a minor merger would help explain the increase in mass at that time, and in both the merger and interaction scenarios, an episode of enhanced star formation may also be triggered, further contributing to the stellar mass of the component.

\section{Discussion}\label{sec:discussion}
\paperI\. of the BUDDI-MaNGA project describes how the galaxies were modelled with a S\'ersic+exponential light profile using \textsc{buddi} to cleanly extract the spectra of the bulge and disc respectively. This paper presents the first scientific study for this project, in which we measured and compared the stellar populations and star-formation histories of the bulges and discs of S0s in the sample in order to understand their evolution and thus how their star formation was truncated. The stellar populations were derived using two different methods- luminosity weighted properties were calculated using the line strengths, and both luminosity- and mass-weighted populations were extracted through full spectral fitting with \textsc{ppxf}. Together these analyses give an overview of both the most recent episodes of star formation in the bulges and discs, and their mass assembly histories.

\subsection{General properties of bulges and discs}\label{sec:bulges}
The first step in the analysis was to derive the luminosity-weighted ages and metallicities of the bulges and discs through interpolation of the measurements of their line strengths from the SSP models. It was found that the bulges tend to lie mostly towards the old, metal-rich region of the model. A trend can be seen with the galaxy mass, such that the bulges in more massive galaxies have generally higher values for their luminosity-weighted age and metallicity, while those in lower-mass galaxies contain younger and more metal-poor stellar populations. The discs on the other hand show a distribution more towards younger stellar populations, with a weaker trend that discs in lower mass galaxies have lower metallicities.  A similar trend was seen in \citet{Fraser_2018b} for both the bulges and discs in their sample of S0s, though in that case they found two clear populations of each component- galaxies with masses  $>10^{10}~\text{M}_\odot$ tend to contain older, more metal-rich populations while lower mass galaxies ($\text{M} < 10^{10}~\text{M}_\odot$) contain younger, more metal-poor populations.  While this work found  trends in age and metallicity with the total mass of the galaxy, such that bulges and discs of higher mass galaxies tend to show older, more metal-rich stellar populations, we find no evidence of two clear populations of S0s. There are a few possible reasons for this discrepancy. For example, in \citet{Fraser_2018b} the bulge and disc spectra were taken from regions where they determined that the light was dominated by that from the bulge and disc respectively. With such a technique, it is still likely that the spectra for each component contains some contamination from the light of the other component. However, the authors quantified the level of contamination using the light profile fits from \citet{Simard_2011} and concluded that any effect should be minimal. It is therefore likely that this discrepancy instead comes from the smaller sample of galaxies used in this study (78 compared to 279) and the mass range of these galaxies. The MaNGA IFUs are allocated to galaxies to ensure coverage of up to 1.5 or 2.5~$R_e$, and so lower mass galaxies were generally observed with the smaller IFUs. Since this work was limited to galaxies observed with the two largest IFUs, the sample is more biased towards higher mass systems than \citet{Fraser_2018b}, leaving the low-mass regime to be more sparsely sampled. A similar conclusion was reached by \citet{Barsanti_2021}, who found no clear evidence of two different populations based on the galaxy mass for their sample of galaxies with masses $>10^{10}M_\odot$.

The line strengths were then used to derive estimates of the $\alpha$-enhancements of the bulges and discs through the Mgb/$\langle$Fe$\rangle$ line ratio. This analysis revealed that the $\alpha$-enhancements of bulges and discs in the same galaxy are correlated. A similar trend was found for S0s in \citet{Johnston_2014} and \citet{Fraser_2018b}. \citet{Johnston_2014} found that the discs had generally slightly higher $\alpha$-enhancements than the bulges, indicating that the discs formed over a shorter star-formation timescale than the bulges. The authors attributed this trend to a scenario where the final episode of star formation in the bulge was fuelled by gas enriched by star-formation in the disc, and in which the star formation in the bulge continued for a short time after being truncated in the disc, potentially through a gas stripping mechanism. This trend however is not observed in Fig.~\ref{fig:stellar_pops_alpha}, where instead there appears to be a weak trend towards the bulges having slightly higher $\alpha$-enhancements than the discs. This difference could be due to either differences in the mass ranges in the two studies, the environments studied, or the methods used, where \citet{Johnston_2014} used S0s with masses of $<10^{10.6}M_\odot$ in the Virgo and Fornax clusters, and applied a 1D bulge-disc decomposition to long-slit spectra from the major axes of their sample.

 \citet{Fraser_2018b} observed that the low-mass galaxies show a larger spread in the bulge and disc $\alpha$-enhancements than the higher mass galaxies, which they attributed to the low-mass galaxies having more varied star-formation timescales as a result of them being more significantly affected by their local environments. While the numbers of galaxies in this study are much smaller, especially in the low-mass regime, it can be seen in Fig.~\ref{fig:stellar_pops_alpha}  that the lower-mass galaxies do show slightly larger scatter than their higher-mass counterparts.

The mass- and luminosity-weighted stellar populations were then measured through full spectral fitting. Both methods revealed that the bulges are generally older and more metal rich than their discs, with the mass-weighted values showing older ages in both components and higher metallicities in the bulges when compared to the luminosity-weighted properties. Since the light from a galaxy, or part of a galaxy, is dominated by the most recent episode of star formation, a recent short period of enhanced star formation can skew the luminosity-weighted ages to younger values, even if the mass created during this episode only accounts for a tiny fraction of the total mass of the galaxy. This scenario is sometimes referred to as `frosting' \citep{Trager_2000}. Similarly, the luminosity-weighted metallicity would reflect that of the gas that fuelled the most recent episode of star formation, which may have originated from within the galaxy or have been accreted from outside. The mass-weighted stellar populations, however, would instead reflect the properties of the majority of the mass within the galaxy. For example, the mean mass-weighted age would give an indication of when the majority of the mass of the galaxy was assembled, while the metallicity would reflect the metallicity of the gas that created that mass. Consequently, the luminosity-weighted ages and metallicities being lower than the mass-weighted values indicates that both components have undergone extended star-formation histories during their lifetimes, with at least some recent star formation occurring since the majority of the mass was built up. Similar results have been seen in previous works. For example, \citet{Johnston_2021} also found evidence of younger luminosity-weighted ages in the bulges and discs than their mass-weighted ages, which they attributed to more recent star formation that contributes towards only a small fraction of the mass of the galaxy. Simulations by \citet{Pfeffer_2022} also found evidence of negative age gradients across cluster galaxies, which they attributed to outside-in quenching of star formation as the galaxies fall into the clusters.

Through all three analyses, a trend can be seen between the ages, metallicities and the total mass of the galaxy. For example, lower-mass galaxies tend to contain younger and more metal-poor stellar populations in their bulges than their higher mass counterparts. This result indicates that the majority of the mass of the bulges formed over a short timescale, with bulges in more massive galaxies forming longer ago. This rapid, early formation of bulges in higher mass galaxies is similar to the monolithic collapse scenario first proposed by \citet{Eggen_1962}, while the slower growth of the bulges in lower mass galaxies could indicate that they formed through secular evolution processes \citep{Gadotti_2001}. The discs show a correlation with metallicity only, such that lower mass galaxies tend to show lower metallicities in their discs, while higher-mass galaxies show a wider range in disc metallicities.  \citet{Pak_2021} found a tight correlation between the luminosity-weighted ages and metallicities of the bulges of S0s using a similar method, with a weaker trend seen in the discs. In this study, no clear trends are seen in the discs, but a potential trend can be seen in the luminosity and mass-weighted ages and metallicities of the bulges in Fig.~\ref{fig:stellar_pops_LWMW} (left plots, particularly towards the old, metal-rich region). This trend is more easily seen when the ages are plotted linearly, as in \citet{Pak_2021}, though a clear offset was found between this work and the trend presented in that paper such that the ages presented here for a given metallicity are lower than those presented in \citet{Pak_2021}. This offset is likely due to a combination of the differences in the methods used (regularized vs unregularized fits with \textsc{ppxf}), differences in the models (MILES spectra created with the  BaSTI \citep{Pietrinferni_2004,Pietrinferni_2006} vs Padova \citep{Girardi_2000} isochrones), and decomposition techniques (3D modelling vs bulge/disc dominated regions).

In order to better understand the evolution of the bulges and discs, their star-formation histories were plotted as their mass assembly as a function of lookback time. For both the bulges and discs, it is clear that the majority of their mass built up quickly early in their lifetimes, with a slower mass accretion later in their lives. In general, the bulges show that the majority of their mass was built up much earlier than in the discs, which show a much broader distribution in star-formation histories. This trend is consistent with previous results that the bulges are generally older than the discs, having built up their mass earlier and over a shorter timescale, and with the scenario where the extended star formation activity occurred mainly in the discs in the progenitor spiral galaxies. This result is in agreement with \citet{Mendez_2021}, who found that the bulges formed first, with their properties tied to those of the host galaxy, while discs formed later. They also found evidence that the formation of the disc is driven by the  properties of the bulge they form around, and thus that the evolution of these two components are strongly linked. They concluded that S0s form through an inside-out formation scenario, where bulges form early and undergo little evolution since, but drive the evolution of the galaxy as a whole.

\subsection{Dependence on galaxy mass}\label{sec:environment}

A clear trend was found between the total mass of the galaxy and the mass-assembly history of the bulges: bulges in higher-mass galaxies, typically $M\geq10^{10}M_\odot$, built up the majority of their mass earlier than those in lower-mass galaxies, which show more extended star formation activity, in some cases starting later on. The discs on the other hand show no such correlation with mass, indicating that their star-formation histories are much more complicated, likely being affected more by stochastic episodes of star formation and accretion of gas and through minor mergers.

The earlier formation of the bulges in the high mass galaxies points towards an inside-out formation scenario, where the bulges formed early in the lifetime of the galaxy and have evolved passively since then while the discs have built up their mass more slowly through extended star formation activity. Thus these galaxies typically have older, redder bulges and younger, bluer discs \citep[e.g.][]{Eggen_1962, Fall_1980, vandenBosch_1998, Kepner_1999,Gonzalez_2014,Breda_2020a}. Using data from the SIMBA and IllustrisTNG simulations respectively, \citet{Appleby_2020} and \citet{Nelson_2021} found that AGN feedback is necessary to suppress the central star formation in a galaxy, thus resulting in the older bulges surrounded by younger discs in these high-mass galaxies.

The similar mass-assembly histories for the bulges and discs of lower mass galaxies however indicate that they formed together, with the similar $\alpha$-enhancements suggesting that they formed from the same material. Similar results were seen by \citet{Fraser_2018b}, \citet{Devergne_2020} and \citet{Johnston_2021}. Of the $\sim10\%$ of galaxies in this sample that show significantly younger mass-weighted ages in the bulges compared to their discs, most of these galaxies are in the low-mass regime. This trend reflects that in these cases, the galaxies formed through an outside-in formation scenario, where the disc forms first and the bulge is built up later on. \citet{Graus_2019} proposed that this outside-in formation scenario could be due to feedback from star formation, which pushes older stars out to larger radii over time, resulting in a positive age gradient. Similarly, the lack of strong black hole feedback in these lower mass galaxies could help drive the outside-in formation scenario since the central star formation is not suppressed \citep{Appleby_2020,Nelson_2021}.

\subsection{Dependence on environment}\label{sec:environment}
All the plots for stellar populations have been presented with the galaxies separated into three environment categories- isolated (43 galaxies with $\leq3$ members), group (25 galaxies with between 4 and 50 members) and cluster (10 galaxies with $>50$ members).  However, since the MaNGA sample wasn't built to probe overdense environments, the galaxies analysed in this study are biased towards less dense environments, and so the results presented in this section should be used with care.

No clear trend was seen between the stellar populations for the bulges and discs and the environment, i.e. the age-metallicity-$\alpha$-enhancement space for the three analysis steps show a wide distribution with all three environment categories. This effect was also seen in \citet{Fraser_2018b} for their sample of MaNGA S0s. Evidence of a dependence on environment has been seen before. For example, using EAGLE simulations \citep{Schaye_2015, Crain_2015}, \citet{Pfeffer_2022} found that cluster S0 galaxies tend to display positive age gradients, which corresponds to outside-in quenching, while more isolated S0 galaxies showed more negative age gradients, reflecting inside-out quenching. \citet{Bedregal_2011} and \citet{Finn_2018} found the same result for galaxies in local clusters. With radial gradients across galaxies, however, it is hard to determine whether they arise from gradients within the disc itself or from the varying bulge/disc light ratio as a function of radius, where the bulge and disc show very different populations. Therefore, the lack of a clear trend between bulge and disc stellar populations as a function of environment could indicate that the gradients come primarily from the quenching of the disc star formation, or simply the low numbers of cluster galaxies used in this study. A photometric study of cluster S0s by \citet{Barsanti_2021b} found that while bulges are generally redder than their surrounding discs, their colours showed no clear correlation with the environment within the cluster, while the discs appeared bluer towards the outskirts of the cluster than in the core. They concluded that in clusters, S0s are primarily formed by processes acting upon the discs within the cluster cores.

The mass assembly plots in Fig.~\ref{fig:SFHs} do show some slight differences for the three environments. The curves for the bulges show very similar trends for all three environments until around 10~billion years ago, where they diverge to reflect the more rapid mass assembly of the bulges in cluster galaxies than in group and isolated galaxies, who built their bulges more slowly since then.  A similar, though less significant, trend can be seen for the discs between 1 and 4 billion years ago. \citet{Coccato_2020} and \citet{Johnston_2021} found that S0s in clusters built up their mass rapidly and have evolved passively since then, while those in more isolated environments are more dispersion supported and showed longer star formation timescales, indicating that they have been more significantly affected by minor mergers. The trends seen in the mass assembly plots in this work agree with this scenario. However the small numbers of galaxies in the group and  cluster samples relative to the isolated sample mean that the results in this study are not particularly robust.  Furthermore, larger differences in the mass assembly curves for the bulges in this work are seen when the sample is split by mass as opposed to environment, indicating that mass is a more important driver for bulge evolution, while disc evolution does not appear to be particularly dependent on the mass or environment of the galaxy. To better understand the significance of this result, a follow-up study of a larger number of bulges and discs will be presented in \textcolor{black}{Jegatheesan et al (in prep)}.

\subsection{Bulges and discs of the same galaxies}\label{sec:bulges_and_discs}
\begin{figure}
 \includegraphics[width=\linewidth]{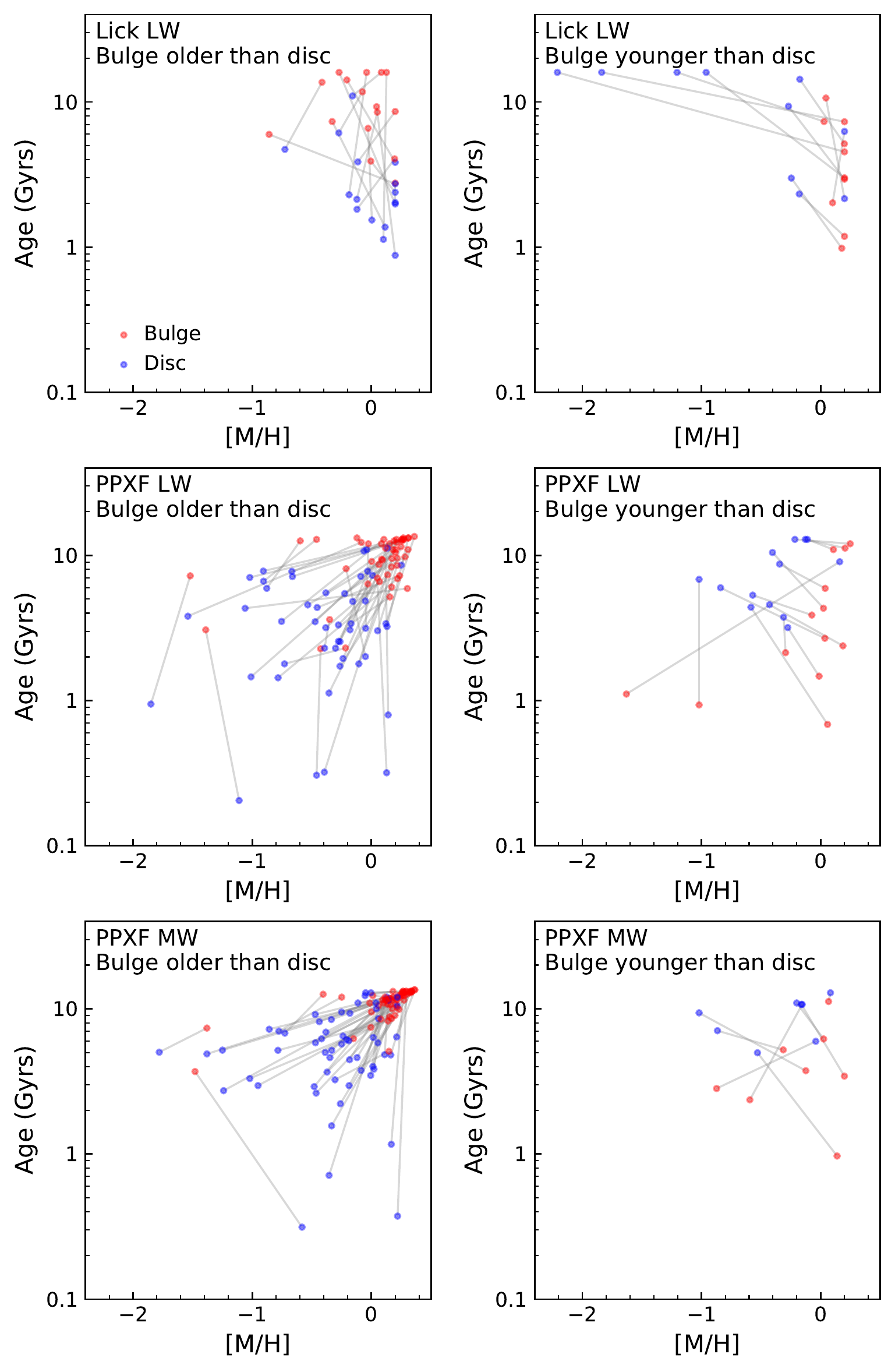}
 \caption{A comparison of the luminosity (top and middle rows) and mass-weighted (bottom row) stellar populations for the bulges and discs using all three methods. The left and right columns show cases where the bulges are significantly older and younger than the discs, respectively, where components are considered to be significantly older or younger if their ages or metallicities are inconsistent within their uncertainties. 
} \label{fig:age_met_comparison}
\end{figure}

Having compared the general properties of bulges and discs in this sample of S0 galaxies, the next step is to compare the bulges and discs in the same galaxy in a more quantitative way to understand how the galaxies evolved. Figure~\ref{fig:age_met_comparison} presents again the luminosity and mass-weighted ages and metallicities for all three methods, this time plotting the bulges and discs together and separating the galaxies into two categories- those with bulges significantly older and significantly younger than the discs. In this case, bulges were determined to be significantly older or younger than the discs if their ages were inconsistent within their uncertainties. The grey lines link the bulges and discs of the same galaxy, and those galaxies that have consistent bulge and disc ages (i.e. within their uncertainties) are not plotted.  While this figure presents no new measurements, its aim is to help demonstrate the significance of the differences in the stellar populations, and to highlight cases where the bulge and disc stellar populations are considered to be truly inconsistent within the same galaxy, thus reflecting their different star formation histories.

One can see immediately that, in general, the majority of the galaxies show older and more metal-rich stellar populations in their bulges than their discs. Comparing this result to Figures~\ref{fig:stellar_pops_BvD}, \ref{fig:stellar_pops_LW} and \ref{fig:stellar_pops_MW} shows that the galaxies that show younger and more metal-poor stellar populations in their bulges tend to be low-mass galaxies, thus suggesting that they have experienced significant amounts of star formation in their bulges more recently than their discs. Table~\ref{tab:age_met_comparison} gives an overview of the number and fraction of galaxies whose bulges contain stellar populations that are older, younger, more metal rich and metal-poor than their discs, and those that have consistent values with their discs, for all three methods used in this paper. It should be noted that not all galaxies were included in Fig.~\ref{fig:age_met_comparison} and Table~\ref{tab:age_met_comparison} for the Lick index method- those galaxies whose line strengths fell outside of the  SSP model grid were assigned the values for the nearest point on the grid, unless their uncertainties also fell outside of the grid, in which case they were removed from the sample. This age trend is similar to that seen in the analysis of S0s in MaNGA by \citet{Dominguez_2020}, who found evidence of a bimodality in the luminosity-weighted stellar populations, where galaxies with masses  $>3\times10^{10}M_\odot$ displayed strong negative age gradients and flat metallicity gradients in galaxies, while galaxies with lower masses tend to display flatter age gradients and stronger negative metallicity gradients. \citet{Tabor_2019} also found evidence that bulges of S0s generally contain more metal-rich stellar populations than discs  while the ages of the two components are comparable.

\begin{table*}
	\centering
	\caption{Differences in the mass- and luminosity-weighted ages and metallicities from the 3 methods, with the mean values at the bottom. For each method, the number of galaxies with successful measurements are given in the second column, and both the number and percentage of galaxies that fall into each criteria are given in the remaining columns.}
	\label{tab:age_met_comparison}
	\begin{tabular}{lrrrrrrrrr} 
		\hline
		  &  &   \multicolumn{3}{c}{Age (Gyrs)}  &   \multicolumn{3}{c}{[M/H]} \\
		  Method  & No. galaxies & B$\gg$D (\%) & B$\ll$D (\%)  & B$\sim$D (\%)  & B$\gg$D (\%) & B$\ll$D (\%) & B$\sim$D (\%)  \\
		\hline
LW (Lick)			& 62	& $18~(29\%)$	& $10~(16\%)$	& $34~(55\%)$ 	& $12~(19\%)$	& $4~(7\%)$	& $46~(74\%)$ 	\\
LW (\textsc{ppxf}) 	& 78	& $49~(63\%)$	& $13~(17\%)$	& $16~(20\%)$ 	& $63~(81\%)$	& $5~(6\%)$	& $10~(13\%)$ 	\\
MW (\textsc{ppxf}) 	& 78	& $58~(75\%)$	& $8~(10\%)$ 	& $12~(15\%)$	& $68~(87\%)$	& $6~(8\%)$	& $4~(5\%)$ 	\\
Mean 		 	&  	& $38~(56\%)$	& $10~(14\%)$ 	& $21~(30\%)$	& $48~(62\%)$	& $5~(7\%)$	& $20~(29\%)$ 	\\
		\hline
	\end{tabular}
\end{table*}

In all three analyses, the bulges were found to generally contain older and more metal-rich stellar populations than their surrounding discs. As can be seen in  Table~\ref{tab:age_met_comparison}, the exact fractions depended on the method used, with the luminosity-weighted ages from the line strengths and from \textsc{ppxf} finding $29\%$ and $63\%$, respectively, of the bulges being significantly older than their discs, while the mass-weighted ages revealed a higher fraction of $75\%$. These results are again consistent with the idea of a recent episode of star formation leading to younger luminosity-weighted ages in the bulges, while the mass-weighted age is unaffected. All three methods reveal only a small fraction of $\sim 14\%$ of bulges containing younger stellar populations than their discs, with  $\sim30\%$ of galaxies showing similar or consistent ages in both components. These numbers show a higher fraction of galaxies with higher mass-weighted ages than \citet{Barsanti_2021}, who found $\sim 23\%$, $\sim 34\%$ and $\sim 43\%$ of galaxies contain significantly older, younger or similar age bulges compared to their discs (note- these numbers are the average of three methods used in that paper). This discrepancy may be due to the different sample sizes (192 vs 78), or the definition of `significantly', where in this paper is simply taken to be outside of the uncertainties (i.e. $1\sigma$) whereas \citet{Barsanti_2021}  use a limit of $3\sigma$. However, recalculating the fractions with this $3\sigma$ limit still shows inconsistencies, with fractions of $\sim 55\%$, $\sim 5\%$ and $\sim 40\%$ of galaxies contain significantly older, younger or similar mass-weighted age bulges compared to their discs. There could also be a mass effect, where  \citet{Barsanti_2021}  uses galaxies of mass $10^{10}-10^{11.5}M_\odot$ while this work extends down to $10^{9}M_\odot$, or an environmental effect, where they use cluster galaxies while this work has a larger fraction of more isolated galaxies. An environmental effect has been found on the bulge and disc star formation histories in S0s in other works. For example, \citet{Coccato_2020} and \citet{Johnston_2021} concluded that the star formation in S0s in clusters is more likely to be truncated through gas stripping processes, while in more isolated galaxies minor mergers play a more significant role. 
It is therefore possible that this extended star-formation timescale in the bulges of cluster galaxies leads to them having lower mass-weighted ages, therefore reducing the difference in the mean bulge and disc ages and thus resulting in a lower fraction of bulges in cluster environments that show older stellar populations than their discs when compared to more isolated S0s. However, to determine whether this scenario is true requires further analysis of a larger sample of galaxies from a wide range of masses and environments, which is beyond the scope of this work. 

As shown in Table~\ref{tab:age_met_comparison}, the relative metallicities of the bulges and discs also appears to depend in the method used, with luminosity-weighted metallicities from line strengths and \textsc{ppxf} showing $19\%$ and $81\%$, respectively, of bulges containing significantly more metal-rich stellar populations than their discs, $7\%$ and $6\%$ showing more metal-poor populations, and $74\%$ and $13\%$ showing no clear difference. On the other hand, the mass-weighted stellar populations show $87\%$ of bulges to be more metal rich, $8\%$ more metal poor, and $5\%$ with no difference. 
Using the $3\sigma$ limit on the metallicity differences results in $\sim 69\%$, $\sim 5\%$ and $\sim 26\%$ of galaxies contain significantly more metal-rich, more metal poor and with similar metallicities. Interestingly, these fractions are consistent with those of \citet{Barsanti_2021}, who determined that the galaxies with significantly more metal rich bulges than discs tended to have redder bulges than discs, and thus concluded that the main driver for the colour of these components is the metallicity. 

As a final note, the results presented above focus on the differences in the stellar populations of the bulges and discs based mainly on the uncertainties in these properties. For a discussion on the impact on the stellar populations due to variations in the fit parameters can be found in \paperI\..

\section{Conclusions}\label{sec:conclusions}
As part of the BUDDI-MaNGA project, we have explored the mass- and luminosity-weighted stellar populations of the bulges and discs in a sample of S0s from the MaNGA survey (DR15) with the aim of understanding their evolution and determining how their star formation was truncated. Our main results are summarized as follows:

\begin{itemize}
 \item In general, bulges contain older and more-metal rich stellar populations than their discs. Analysis of their mass-assembly histories revealed that bulges built up the majority of their masses over short timescales early in their lifetimes, while discs formed around them over longer timescales.
 \item The $\alpha$-enhancements of the bulges and discs in the same galaxy are correlated, indicating that the most recent episodes of star formation in the bulges and discs are somehow connected. Since the luminosity-weighted ages of the bulges are lower than the mass-weighted values, this correlation in the $\alpha$-enhancement indicates that the bulges have likely experienced a `frosting' of star formation since the majority of their mass was formed that was fuelled by gas enriched by star formation in the discs. 
 \item The star-formation histories of bulges appear to be linked to the total mass of the galaxy, such that bulges in lower mass galaxies ($<10^{10}M_\odot$) formed more recently over more extended timescales, while those in higher mass galaxies formed rapidly early in the lifetime of the galaxy. Discs on the other hand show no correlation with mass of the galaxy, indicating that they experienced more diverse star formation histories, likely affected more by bursty star formation, interactions, mergers and accretion of material.
 \item A small but significant fraction of galaxies do show evidence of bulges containing younger and/or more metal poor stellar populations than their discs. The majority of these galaxies have masses $<10^{10}M_\odot$, indicating that this trend may be connected to the more recent formation of the bulges 
 \item While a small difference was seen in the mass assembly histories of the bulges and discs as a function of environment, such that cluster galaxies appear to have built up the majority of their mass earlier than those in group and isolated environments, the effect was very small and could be affected by the low number statistics. It therefore appears that in this sample of galaxies, the galaxy mass plays a more significant role in the evolution of the bulges and discs, and therefore of the galaxy itself.
 \item Our results point towards an inside-out formation scenario for high mass ($M\geq10^{10}M_\odot$) S0s, where the bulges form first over a short timescale and undergo little evolution since then. The discs formed around the bulges, and experience a wider range of star-formation histories, likely due to ongoing star formation, minor mergers and accretion of material from satellite galaxies. When the star formation in these galaxies was quenched, little or no star formation was triggered in the bulge. Lower-mass S0s, on the other hand, show no clear trends in the age differences between their bulges and discs, indicating that they may have formed together. Some of these galaxies do show evidence of younger mass and light-weighted ages in their bulges than their discs, which may indicate either an outside-in formation scenario, or an outside-in quenching such that a final episode of star formation occurred in their bulges as the disc star formation was truncated.
\end{itemize}

Our results imply multiple formation mechanisms for S0 galaxies, the pathway of which is chiefly determined by a galaxy's current stellar mass, and further analysis is required to explore and understand the dependence on the galaxy mass and the local environment. A series of follow-up studies will be carried out on a wider range of morphologies as part of the BUDDI-MaNGA project to better understand the evolution of galaxy bulges and discs, looking at the effects on their stellar populations and physical parameters and what those clues tell us about the formation of the galaxies we see today and their morphological transformations.

\section*{Acknowledgements}
The authors would like to thank the referee for their useful comments, which have helped improve this paper, and Joel Pfeffer for a useful discussion on the mass and environmental dependences in our results.
E.J.J. acknowledges support from FONDECYT Iniciaci\'on en investigaci\'on 2020 Project 11200263. K.J. acknowledges financial support from ANID Doctorado Nacional 2021 project number 21211770. Y.J. acknowledges financial support from FONDECYT Iniciaci\'on 2018 No. 11180558 and ANID BASAL Project FB210003. Y.O.-B. acknowledges support from FONDECYT Postdoctoral Fellowship Project No.3210442. G.G. gratefully acknowledges support by the ANID BASAL projects ACE210002 and FB210003. Finally, we  thank the bomberos de Santiago for extinguishing the fire at PUC before it reached the server room where the BUDDI-MANGA data are stored.

Funding for the Sloan Digital Sky Survey IV has been provided by the Alfred P. Sloan Foundation, the U.S. 
Department of Energy Office of Science, and the Participating Institutions. 

SDSS-IV acknowledges support and resources from the Center for High Performance Computing  at the University of Utah. The SDSS website is www.sdss.org.

SDSS-IV is managed by the Astrophysical Research Consortium for the Participating Institutions of the SDSS Collaboration including the Brazilian Participation Group, the Carnegie Institution for Science, Carnegie Mellon University, Center for Astrophysics | Harvard \& Smithsonian, the Chilean Participation Group, the French Participation Group, Instituto de Astrof\'isica de Canarias, The Johns Hopkins University, Kavli Institute for the Physics and Mathematics of the Universe (IPMU) / University of Tokyo, the Korean Participation Group, Lawrence Berkeley National Laboratory, Leibniz Institut f\"ur Astrophysik Potsdam (AIP),  Max-Planck-Institut f\"ur Astronomie (MPIA Heidelberg), Max-Planck-Institut f\"ur Astrophysik (MPA Garching), Max-Planck-Institut f\"ur Extraterrestrische Physik (MPE), National Astronomical Observatories of China, New Mexico State University, New York University, University of Notre Dame, Observat\'ario Nacional / MCTI, The Ohio State University, Pennsylvania State University, Shanghai Astronomical Observatory, United Kingdom Participation Group, Universidad Nacional Aut\'onoma de M\'exico, University of Arizona, University of Colorado Boulder, University of Oxford, University of Portsmouth, University of Utah, University of Virginia, University of Washington, University of Wisconsin, Vanderbilt University, and Yale University.


\section*{Data Availability}
The MaNGA datacubes used in this analysis are publicly available through the SDSS DR15 and the associated Value Added Catalogs.



\bibliographystyle{mnras}
\bibliography{buddi_manga2.bib} 

%

\bsp	
\label{lastpage}
\end{document}